\documentclass[11pt,a4paper]{article}
\pdfoutput=1

\usepackage{jheppub}

\usepackage{enumerate}
\usepackage{subfigure}
\usepackage{amsmath}
\usepackage{amssymb}
\usepackage{graphicx}
\usepackage{placeins}
\usepackage{xspace,slashed}
\usepackage{hyperref}
\hypersetup{colorlinks=true, citecolor=blue, urlcolor=blue, linkcolor=blue}
\usepackage[normalem]{ulem}
\usepackage{booktabs,bbm}
\usepackage{wrapfig,multirow}

\usepackage{cleveref}
\crefname{table}{Table}{Tables}
\crefname{equation}{Eq.}{Eqs.}
\crefname{appendix}{App.}{Apps.}
\crefname{section}{Sec.}{Secs.}
\crefname{figure}{Fig.}{Figs.}
\usepackage[bottom]{footmisc}

\usepackage{braket,physics}

\newcommand{\lag}{\mathcal{L}}

\newcommand{\expv}[1]{\langle #1 \rangle}

\title{
   BSM patterns in scalar-sector coupling modifiers
}
\abstract{We consider what multiple Higgs interactions may yet reveal about the scalar sector. We estimate the sensitivity of a Feynman topology-templated analysis of weak boson fusion Higgs pair production at present and future colliders --- where the signal is a function of the Higgs coupling modifiers $\kappa_V$, $\kappa_{2V}$, and $\kappa_\lambda$.
While measurements are statistically limited at the LHC, they are under general perturbative control at present and future colliders, departures 
from the SM expectation give rise to a significant future potential for BSM discrimination in $\kappa_{2V}$. We explore the landscape of BSM models in the space of deviations in $\kappa_V$, $\kappa_{2V}$, and $\kappa_\lambda$, highlighting models that have measurable order-of-magnitude enhancements in either $\kappa_{2V}$ or $\kappa_\lambda$, relative to their deviation in the single Higgs coupling $\kappa_V$.
}
\author[a]{Christoph Englert,} 
\author[a]{Wrishik Naskar,}
\author[a]{Dave Sutherland}
\affiliation[a]{School of Physics \& Astronomy, University of Glasgow, Glasgow G12 8QQ, United Kingdom}
\emailAdd{christoph.englert@glasgow.ac.uk}
\emailAdd{w.naskar.1@research.gla.ac.uk}
\emailAdd{david.w.sutherland@glasgow.ac.uk}

\begin{document}
\maketitle
\allowdisplaybreaks
\section{Introduction}
\label{sec:intro}
The so-called $\kappa$ framework that has been put forward in parallel to the Higgs boson discovery~\cite{LHCHiggsCrossSectionWorkingGroup:2011wcg} 
to assist its characterisation programme has proved a helpful tool in gaining a qualitative understanding of Higgs boson physics. The modifier for a coupling $g$ present
in the Standard Model (SM) $g=g_{\text{SM}}$ is defined as 
\begin{equation}
\label{eq:kappadef}
\kappa_g = {g\over g_{\text{SM}}}\,.
\end{equation}
The $\kappa$ framework traces modifications of the Lorentz structures present in the SM exclusively. This leads to violation of gauge invariance with detrimental implications, which signifies the need to consider a more
flexible theoretical setting. A comprehensive
approach based on effective field theory (in its linear or non-linear realisations) elevates this programme to a better-grounded framework.
Theoretical consistency plays an important role when more data becomes available at the Large Hadron Collider (LHC), thus enabling and necessitating a more 
detailed study of Higgs properties beyond tree level.
 Yet, practical considerations related to the LHC's sensitivity to certain coupling
modifications leave the $\kappa$ framework still applicable in a range of processes. One such process is weak boson fusion (WBF) Higgs pair production, where
the $\kappa$ framework sees continued application. In, \emph{e.g.}, the recent~\cite{ATLAS:2023qzf}, modifications of the (trilinear) Higgs self-coupling $\kappa_\lambda$ and the quartic gauge-Higgs couplings $\kappa_{2V}$ have been constrained
\begin{equation}
\kappa_\lambda  \in [-3.5,11.3]\,,~\quad~\kappa_{2V} \in  [0.0,2.1]\,,
\end{equation}
with similar sensitivity in other di-Higgs final state channels, \emph{e.g.}~\cite{ATLAS:2015sxd,CMS:2017hea,CMS:2017yfv,CMS:2020tkr,ATLAS:2021ifb,CMS:2022hgz,ATLAS:2022xzm}. The $\kappa$ framework also successfully captures the dominant source
of coupling modifications in many concrete UV extensions. 

The Higgs self-coupling corresponds to a unique operator in the dimension 6 effective field theory expansion~\cite{Grzadkowski:2010es}. Modifications of $\kappa_\lambda\neq 1$ can therefore be housed theoretically consistently, which is also demonstrated by $\kappa_\lambda$ investigations
beyond tree level~\cite{McCullough:2013rea,Degrassi:2016wml,Gorbahn:2016uoy,Kribs:2017znd} that do not lead to theoretical inconsistencies.

\begin{figure*}[!t]
\centering
\includegraphics[width=0.95\textwidth]{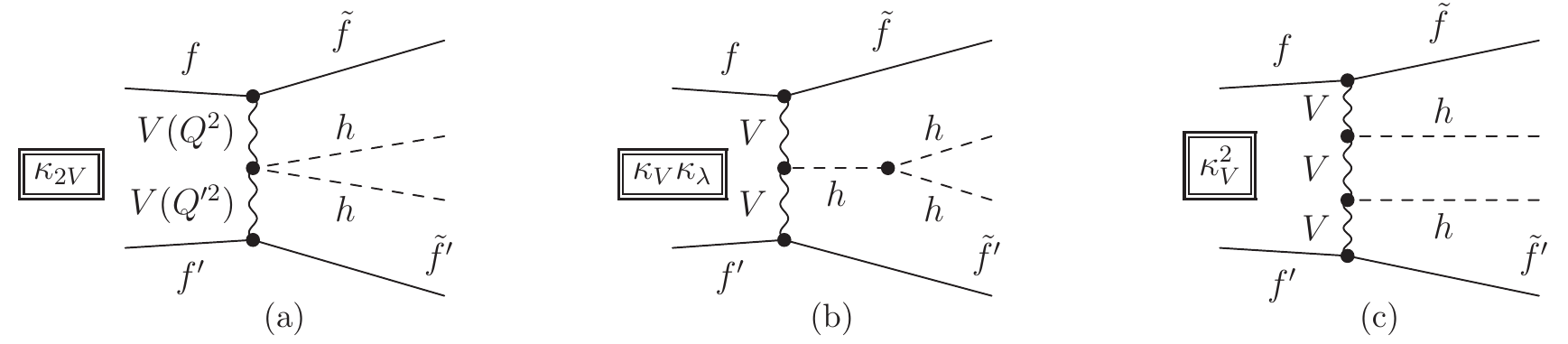}
\caption{Representative Feynman diagram topologies contributing to $\kappa_{2V}$ sensitivity via weak boson fusion. \label{fig:diags}}
\end{figure*}

This is vastly different for $\kappa_{2V}\neq 1$, which breaks electroweak gauge invariance in the SM(EFT), leading to a breakdown of renormalisability.
As $\kappa_{2V}$ is related to a modification of the gauge sector, a departure from $\kappa_{2V}=1$ requires care when moving beyond tree-level considerations~\cite{Anisha:2022ctm}. Notwithstanding these theoretical obstacles, the gauge-Higgs quartic interactions can be strong indicators of Higgs compositeness as a consequence of dynamical vacuum misalignment. For instance, in minimal theories of Higgs compositeness~\cite{Agashe:2004rs,Contino:2006qr}, the typical deviations in the Higgs-gauge sector are given by
\begin{equation}
\label{eq:relation}
\kappa_{V} = \sqrt{1 - \xi}\,, \quad \kappa_{2V} = 1 - 2\xi\,,
\end{equation}
where $\xi$ measures the electroweak vacuum expectation value $v\simeq 246~\text{GeV}$ in units of the CCWZ~\cite{Coleman:1969sm,Callan:1969sn} order parameter. These relations are independent of the mechanism responsible for vacuum misalignment; they are independent of how partial compositeness is included in the fermion sector.
In contrast, for scenarios
of iso-singlet mixing~\cite{Binoth:1996au,Patt:2006fw,Schabinger:2005ei,Englert:2011yb}, we obtain
\begin{equation}
\kappa_{V} = \cos\alpha\,, \quad \kappa_{2V} = \cos^2\alpha\,.
\end{equation}
It is immediately clear that a sufficiently precise measurement of the quartic gauge-Higgs coupling, $\kappa_{2V}$, serves as a discriminator between these two dramatically different BSM scenarios. While it is always possible to interpret $\kappa_V<1$ in either scenario through identifying $\cos\alpha = \sqrt{1-\xi}$, 
\begin{equation}
\cos 2\alpha \neq \cos^2\alpha
\end{equation}
away from the decoupling limit $\alpha\neq 0 \neq \xi$ in either BSM scenario.

In this note, we provide a detailed discussion about the relevance of $\kappa_{2V}$ and its relation to $\kappa_V$, $\kappa_\lambda$ for BSM physics, spanning from traditional renormalisable scenarios to effective theories. We will focus on the phenomenology in WBF di-Higgs production at the LHC and its extrapolation to other future collider environments, where these three couplings enter on an a priori equal footing.
Including a viewpoint of geometry~\cite{Alonso:2015fsp}, we clarify the sensitivity to BSM scenarios that can be gained in a range of models (and their deformations) at the high-luminosity (HL-)LHC phase, as well as future colliders such as the FCC-hh or electron-positron machines. In passing, we include a discussion of higher-order QCD effects at hadron machines. 

This work is organised as follows: In Sec.~\ref{sec:present}, we discuss the future sensitivity that can be achieved in a Feynman-diagram templated analysis as performed by the experiments, extrapolating from the current sensitivity that they report. This forms the phenomenological backdrop of the theoretical reach of this sensitivity in Sec.~\ref{sec:theory}. There we categorise the $\kappa_V$, $\kappa_{2V}$, $\kappa_\lambda$ parameter space in terms of renormalisable theories and effective theories of compositeness and their deformations. We summarise in Sec.~\ref{sec:conc}, which provides an outlook towards monetising the BSM potential of $\kappa_{2V}$ sensitivity in these experimental environments in relation to $\kappa_V$ and $\kappa_\lambda$.

\section{Present and future of scalar couplings in WBF Higgs pair production}
\label{sec:present}

In this section, we consider the experimental limit on $\kappa_V$, $\kappa_{2V}$, and $\kappa_\lambda$, as defined by the tree-level lagrangian
\begin{equation}
  \lag = \frac12 (\partial h)^2 +m_W^2  \left( W^+_\mu W^{-\mu} + \frac{1}{2c_W^2} Z_\mu Z^{\mu} \right) \left[ 1 + \kappa_V \, \frac{2h}{v} + \kappa_{2V} \, \frac{h^2}{v^2} \right]  - \kappa_\lambda \, {m_h^2\over 2 v} h^3 \, ,
  \label{eq:kappaVand2Vdef}
\end{equation}
with $v=246~\mathrm{GeV}$ and cosine of the Weinberg angle $c_W=\cos\theta_W$. Tree-level custodial symmetry is assumed throughout (\emph{i.e.}, $\kappa_W=\kappa_Z=\kappa_V$ and $\kappa_{2W}=\kappa_{2Z}=\kappa_{2V}$).

Before we turn to the relevance of the different couplings and their BSM reach, it is instructive to clarify the anticipated shape of the $\kappa_V-\kappa_{2V}$ exclusion. WBF probes the incoming weak bosons at space-like momenta in the diagrams of Fig.~\ref{fig:diags}.
However, the analysis of $Wh \to Wh$ scattering for physical momenta (the crossed process that enters WBF subdiagrams of \cref{fig:diags}) provides insight into gauge-symmetry cancellations that carry over into qualitative phenomenological outcomes via the effective $W$ approximation~\cite{Dawson:1984gx,Contino:2010mh}. In the high energy limit $\sqrt{s}\gg m_h+m_W$, the polarised amplitudes scale as\footnote{This result can be straightforwardly obtained with FeynArts/FormCalc/LoopTools~\cite{Hahn:1998yk,Hahn:2000jm,Hahn:2000kx}, which we use throughout this work.
Phase-space and polynomial suppression of the valence quark parton distribution functions significantly modify these naive expectations.}
\begin{equation}
\label{eq:highenergy}
{m^2_W\over s} {\cal{A}}(W_L h\to W_L h)
 \sim  {m_W\over \sqrt{s}}  {\cal{A}}(W_T h\to W_L h) 
  \sim \kappa_{2V}-\kappa_{V}^2  \,.
\end{equation}
This shows that in the SM (as well as for direct Higgs mixing), we can expect a significant destructive interference to maintain unitarity at high energies. The scaling with energy in the WBF process is pdf-suppressed for massless partons (including in the effective $W$ approximation), however, large enough deviations from the SM correlation manifest themselves as an enhanced cross section so that
limits can be set. Note that $\kappa_\lambda$ does not enter the $Wh$ amplitude with energy enhancement, and its constraints are therefore set by a priori perturbativity limits (see~\cite{DiLuzio:2017tfn,Arco:2020ucn,Arco:2022lai} for more model-specific considerations).

\begin{figure*}[!t]
\centering
\subfigure[]{\includegraphics[width=0.48\textwidth]{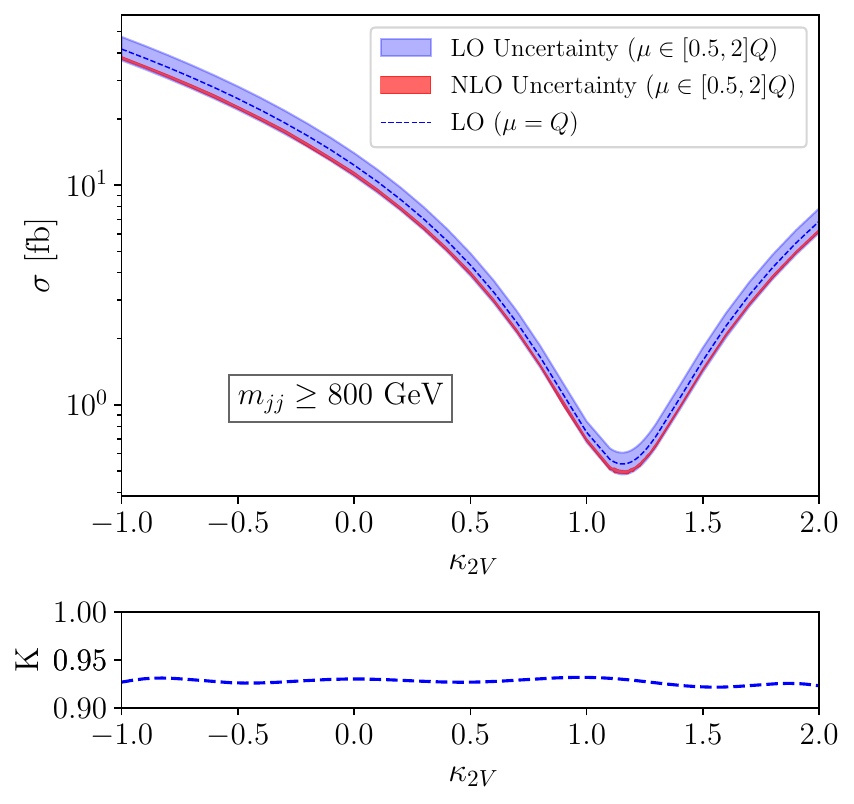}}\hfill
\subfigure[]{\includegraphics[width=0.48\textwidth]{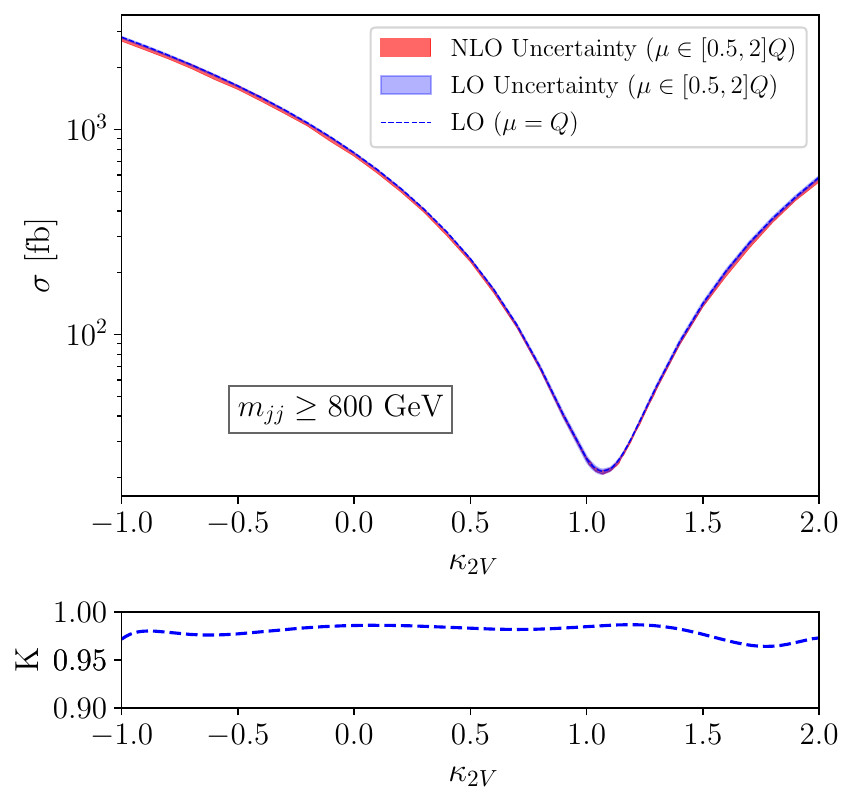}}
\caption{\label{fig:lo_nlo} Next-to-leading order QCD corrections to WBF $pp\to hh jj$ production in the large invariant tagging jet mass region $m_{jj}\geq 800~\text{GeV}$ at the LHC (a) and FCC-hh (b) as a function of $\kappa_{2V}$ for all other parameters chosen to be SM-like.}
\end{figure*}

\subsection{Hadron collider constraints on $\kappa_{V}$ and $\kappa_{2V}$\label{subsec:pp}}
$pp\to hhjj$ production follows $hjj$ production from the point of view of QCD. Therefore, the search region that is selected by the LHC experiments exploits the usual WBF paradigm~\cite{Cahn:1983ip,Rainwater:1998kj,Zeppenfeld:2000td,Figy:2003nv}. Similar to the findings in single-Higgs and double-Higgs production via WBF~\cite{Figy:2008zd,Dreyer:2018qbw}, the QCD corrections can be formidably captured through an adapted choice of the renormalisation and factorisation scales, due to the process' `double-Deep Inelastic Scattering' structure. In particular, in the $\kappa_{2V}$ measurement region selected by the LHC experiments then becomes extraordinarily stable: In \cref{fig:lo_nlo} we show the $pp\to hh jj$ WBF cross section component at next-to-leading order QCD in comparison to the leading order estimate, employing the central scale choice $\mu=Q$ (\emph{cf.} \cref{fig:diags}) for the signal region characterised by an invariant jet mass of $m_{jj}\geq 800~\text{GeV}$.\footnote{We obtain this through a modification of the publicly available {\sc{Vbfnlo}} Monte Carlo programme~\cite{Figy:2003nv,Arnold:2008rz,Baglio:2012np}. Modifications have been cross-checked against \textsc{MadGraph}\_aMC@NLO~\cite{Alwall:2014hca}. See also~\cite{Frederix:2014hta} for studies of rescalings of the Higgs self-coupling.} This extends to the FCC-hh at a significantly increased cross-section, \cref{fig:lo_nlo}.
The cross-section is well-approximated by the LO $\mu=Q$ choice, and the NLO corrections are modest $\sim 5\%$ at the LHC,  decreasing in relevance at the FCC-hh, running at 100 TeV~\cite{FCC:2018vvp}. The key limitation of setting constraints on $\kappa_{2V}$ is then statistics and background systematics.

The ATLAS Collaboration has previously published search results for nonresonant $hh\to b \bar b b \bar b $ production using $27~\text{fb}^{-1}$ of early Run 2 data~\cite{ATLAS:2018rnh}, as well as a dedicated search for VBF $hh$ production in $126~\text{fb}^{-1}$ of data collected between 2016 and 2018~\cite{ATLAS:2020jgy}. The latest analysis~\cite{ATLAS:2023qzf} builds upon these earlier results by incorporating the 2016--2018 data for both production channels and taking advantage of improvements in jet reconstruction and $b$-tagging techniques. Notably, the analysis employs an entirely data-driven technique for background estimation, utilising an artificial neural network to perform kinematic reweighting of data to model the background in the region of interest, and it currently restricts $\kappa_{2V}$ to
\begin{equation}
  \kappa_{2V} \in [-0.0, 2.1] ~([-0.1, 2.1]),~\text{ observed (expected) $95\%$ CL.}
  \label{eq:ATLASbound}
\end{equation}
The CMS Collaboration has also published results of a search for nonresonant $hh\rightarrow b \bar b b \bar b $ with its full Run 2 dataset \cite{CMS:2022cpr}, which restricts the allowed interval for $\kappa_\lambda$ to $[-2.3, 9.4]$ $([-5.0, 12.0])$, at $95\%$ confidence level (CL). Furthermore, a more recent CMS publication~\cite{CMS:2022nmn} that exploits topologies arising from highly energetic Higgs boson decays into $b\bar{b}$, restricts the allowed interval for $\kappa_{2V}$ to
\begin{equation}
  \kappa_{2V} \in [0.62, 1.41]~([0.66, 1.37]),~\text{at observed (expected) $95\%$ CL.}
  \label{eq:CMSbound}
\end{equation}
 ATLAS and CMS have conducted investigations into non-resonant $hh$ in the $ b\bar b \tau^+ \tau^-$~\cite{ATLAS:2022xzm,CMS:2017hea,CMS:2017yfv,CMS:2022hgz,ATLAS:2015sxd} and $ b\bar b \gamma \gamma$~\cite{ATLAS:2021ifb,CMS:2020tkr,ATLAS:2015sxd} decay channels as well. In our analysis, we look at final state topologies for the highest di-Higgs branching ratios, \emph{i.e.}, 
\begin{subequations}
\begin{alignat}{3}
p p \to hh j j \to b \bar{b} b \bar{b} j j ,
\end{alignat}
and,
\begin{alignat}{3}
p p \to hh j j \to b \bar{b} \tau^+ \tau^- j j.
\end{alignat} 
\end{subequations}
We generate events for each process scanning over the space of $\kappa_V$ and $\kappa_{2V}$, keeping $\kappa_\lambda=1$ fixed, and perform a $\chi^2$ fit to obtain the limits on the corresponding couplings.

\subsubsection*{Event generation and selection}
To investigate WBF processes, we use  \textsc{MadGraph}\_aMC@NLO~\cite{Alwall:2014hca} with leading order precision at $13~\text{TeV}$ to generate our events and apply stringent cuts at the generator level on the WBF jet pair's invariant mass ($m_{jj}>800$ GeV). Additionally, we set the pseudorapidity of the $b$-jets to be $|\eta_b|<2.5$ to ensure that the $b$-jets produced from the Higgs pair are centrally located. We modify the \textsc{MadGraph} source code to include the $\kappa$ modifiers for event generation. 
Subsequently, we shower the events with \textsc{Pythia} 8.3~\cite{Bierlich:2022pfr} and use~\textsc{MadAnalysis}~\cite{Conte:2012fm} that interfaces {\sc{FastJet}}~\cite{Cacciari:2011ma,Cacciari:2005hq}, to reconstruct the final state particles with a $70\%$ $b$-tagging efficiency.

Firstly, we define forward and central jets with the following selection criteria:
\begin{enumerate}
\item Forward jets: $p^j_T>30~\text{GeV}$ and $2.5<|\eta_j|<4.5$.
\item Central jets: $p^j_T>40~\text{GeV}$ and $|\eta_j|<2.5$.
\end{enumerate} 
To select events in the $hh \to b \bar b b \bar b$ channel, we utilise the methodology described in~\cite{ATLAS:2023qzf}. Initially, we identify the jet pair with the highest invariant mass as the WBF jets and impose the forward-jet criteria on them. Next, we require at least 4 centrally located jets, all of which must be $b$-tagged. We also apply an additional pseudorapidity separation cut of $|\eta_{jj}|>3$ and an invariant mass cut of $m_{jj}>1~\text{TeV}$ on the WBF-jets. To isolate the WBF region, we further demand that the transverse component of the momentum vector sum of the two WBF jets and the four jets forming the Higgs boson candidates be less than $65~\text{GeV}$. The Higgs pair is constructed from the 4 $b$-jets and a minimum invariant mass of $M_{hh}>400~\text{GeV}$ is required.

For the $hh \to b \bar b \tau^+ \tau^-$ channel, the WBF jet pair is chosen in the same manner as before, and we apply the same cuts on them. However, only two centrally located $b$-tagged jets are required in this case. The $\tau$-leptons can decay either hadronically or leptonically, with the latter being selected using the criteria outlined in the latest ATLAS analysis~\cite{ATLAS:2022xzm}, with a minimum $p_T$ of $15~\text{GeV}$ and limited to $|\eta_l|<2.47$. The light jets arising from the hadronic decay of the $\tau$-leptons are selected with a minimum $p_T$ of $10~\text{GeV}$ and $|\eta_j|<2.5$. Additionally, the leptonic decay of the $\tau$'s results in missing energy. The di-Higgs invariant mass is then constructed using the two $b$-jets and the $\tau$-decay products, with a minimum invariant mass requirement of $M_{hh}>400~\text{GeV}$.

\subsubsection*{Sensitivity and projections\label{subsub:statan}}

We use the distribution of the reconstructed kinematic observable $M_{hh}$ to obtain the current and projected limits on the respective couplings. The $\chi^2$ statistic for our analysis is computed as
\begin{subequations}
\begin{equation}
\chi^2 (\kappa_V,\kappa_{2V})=(b^i_{\text{BSM}}(\kappa_V,\kappa_{2V})-b^i_{\text{SM}})V^{-1}_{ij}(b^j_{\text{BSM}}(\kappa_V,\kappa_{2V})-b^j_{\text{SM}}),
\end{equation}
where $b^i_{\text{BSM}}(\kappa_V,\kappa_{2V})$ represents the combined number of events in the $i$th bin of $M_{hh}$ from both decay channel, considering their respective cross-sections and efficiency, at a given luminosity for a particular value of $\kappa_V$ and $\kappa_{2V}$, and $b^i_{\text{SM}}$ corresponds to the expected number of events solely from the SM for $\kappa_V=1$ and $\kappa_{2V}=1$. 
The covariance matrix $V_{ij}$ is the sum-in-quadrature of two terms: 1) the statistical uncertainties computed from the root of bin entries, 
\emph{i.e.}, the Poisson uncertainty associated with each bin, $b^i_{\text{SM}}$, and 2) fully correlated relative fractional uncertainties ($\varepsilon_{\text{rel.}}$), \emph{i.e.},
\begin{equation}
V_{ij}=b^i_\text{SM} \delta_{ij}+\varepsilon^2_{\text{rel.}}b^i_{\text{SM}}b^j_{\text{SM}} \,.
\end{equation}
\end{subequations}
To fix $\varepsilon_{\text{rel.}}$, we set $\kappa_V=1$ and scan over $\kappa_{2V}$ such that we reproduce a 95\% CL limit of $\kappa_{2V} \in [0.2,2.0]$ at $126 \, \mathrm{fb}^{-1}$ at the LHC, comparable to the limits set by ATLAS~\cite{ATLAS:2023qzf} and CMS~\cite{CMS:2022nmn}. 

\begin{figure*}[!t]
\centering
\subfigure[]{\includegraphics[width=0.48\textwidth]{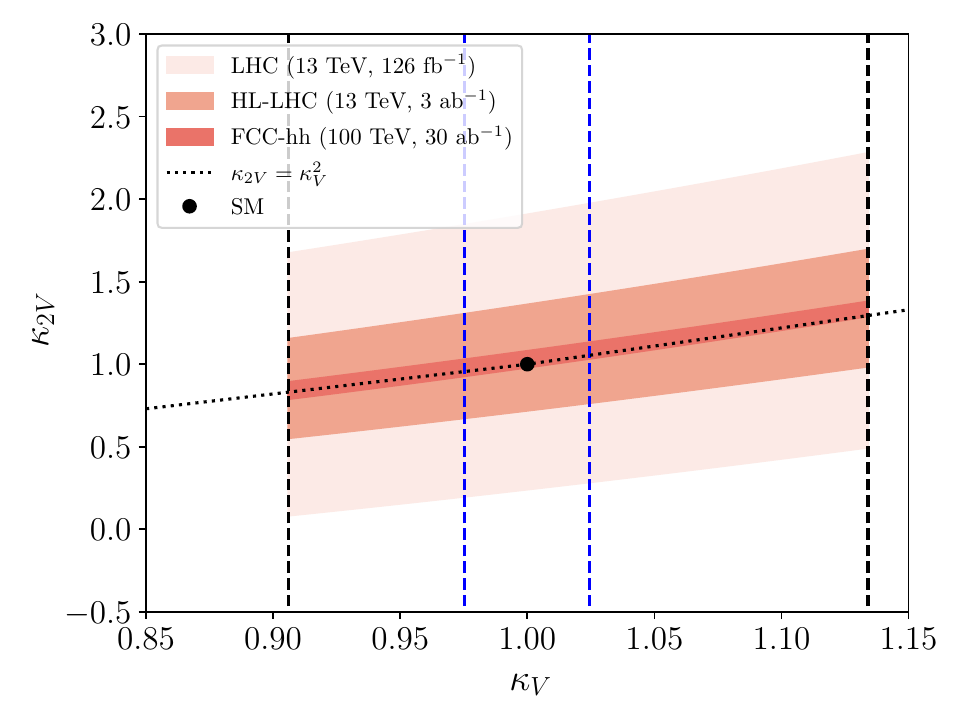}}\hfill
\subfigure[]{\includegraphics[width=0.48\textwidth]{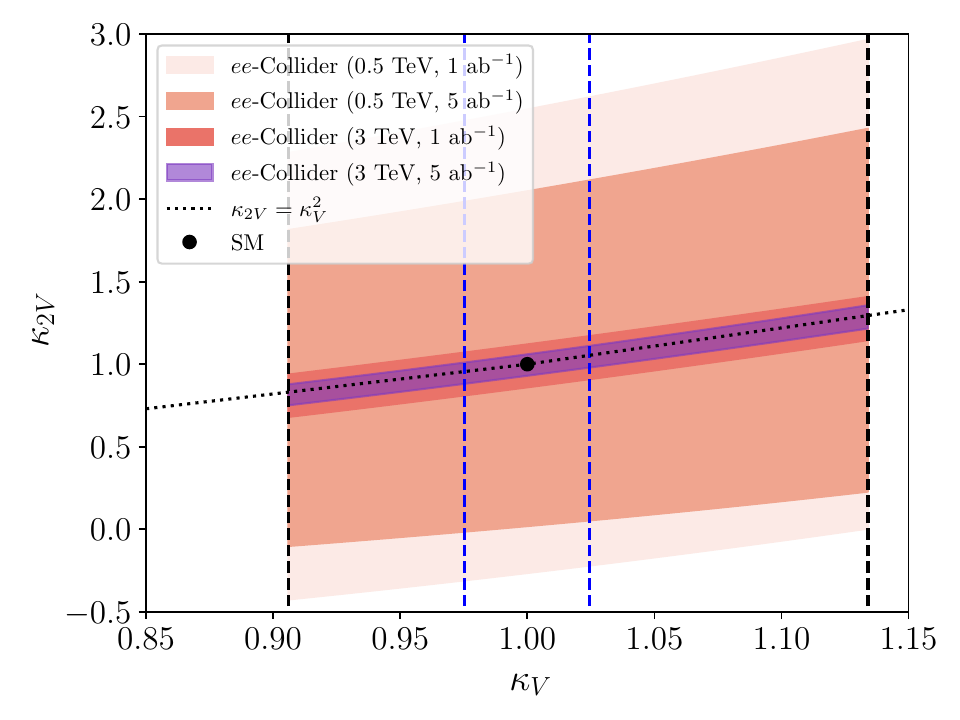}}
\caption{\label{fig:collider} Current and projected constraints on $\kappa_V$ and $\kappa_{2V}$ for various colliders. The current LHC and projected HL-LHC limits on $\kappa_V$, represented by the black and blue dashed lines respectively, have been set using Higgs data from ATLAS~\cite{ATLAS:2022vkf}.}
\end{figure*}

Once our methodology was validated through a comparison of our $\kappa_{2V}$ constraints with the predictions of ATLAS (\cref{eq:ATLASbound}) and CMS (\cref{eq:CMSbound}), we proceed to explore the parameter space of $\kappa_V$ and $\kappa_{2V}$ and obtain constraints at the $95\%$ CL. Our results are presented in \cref{fig:collider}, where we show the $95\%$ confidence bands for the LHC with an integrated luminosity of $126~\text{fb}^{-1}$, the constraints for the High-Luminosity (HL-LHC) frontier with an integrated luminosity of $3~\text{ab}^{-1}$, as well as the projected constraints for the Future Circular Collider (FCC-hh) with $\sqrt{s}=100~\text{TeV}$ and an integrated luminosity of $30~\text{ab}^{-1}$, assuming the same $\varepsilon_{\text{rel.}}$ as for the LHC case. It should be added that using the $\chi^2$ calibrated to \cite{ATLAS:2023qzf,CMS:2022nmn} is a conservative extrapolation for the FCC-hh. A limiting factor in this environment is the reduction of QCD multi-jet contributions as central jet vetoes are not available to suppress these efficiently. This can lead to a considerable variation of the expected sensitivity~\cite{Dolan:2013rja,Dolan:2015zja,Bishara:2016kjn}, in particular when considering the rejection of the irreducible gluon fusion component.

\subsection{Lepton collider constraints on $\kappa_{V}$ and $\kappa_{2V}$ \label{subsec:ee}}
The di-Higgs sector exploration at upcoming lepton colliders has garnered significant attention due to their exceptional sensitivity range, which is attributed to significantly lower background interference compared to hadron colliders. The works of \cite{Chacko:2017xpd,DiVita:2017vrr, Li:2017daq, Abramowicz:2016zbo} (see also Refs.~\cite{Domenech:2022uud,Gonzalez-Lopez:2020lpd}) have been instrumental in investigating the di-Higgs sector and obtaining exclusion limits on $\kappa_\lambda$ and $\kappa_V$. Building on this success, we extend the scope of this exploration by attempting to obtain limits on $\kappa_V$ and $\kappa_{2V}$ for $e^+ e^-$-colliders, adopting a methodology similar to that presented in Sec. \ref{subsec:pp}. In this scenario, the di-Higgs decay into four $b$-quarks is very attractive since the background is orders of magnitude smaller, making it the primary focus of our analysis. The process we want to look at is, therefore,
\begin{equation}
e^+ e^- \to h h e^+ e^- / \nu_e \bar \nu_e \to b \bar b  b \bar b e^+ e^- / \nu_e \bar \nu_e.
\end{equation}
The dominant contributions to the production cross-section come from the WBF process and Higgs-strahlung. We generate our events with \textsc{MadGraph} for two benchmark collider beam energies of $0.5~\text{TeV}$ and $3~\text{TeV}$. To select our WBF signal region, we closely follow the analysis in~\cite{Chacko:2017xpd}. We firstly impose a strict missing energy cut greater than $30~\text{GeV}$. Our study further requires four centrally located $b$-tagged jets, with $p_T>20~\text{GeV}$ and $|\eta|<2.5$. In order to reconstruct the Higgses individually from the $b$'s, we determine the labelling of the four $b$-jets, $b_i$, that minimises the quantity $(m_{12}/\text{GeV}-125)^2+(m_{34}/\text{GeV}-125)^2$, where $m_{ij}$ represents the invariant mass formed by jets $b_i$ and $b_j$. We infer that $b_1$, $b_2$ come from one of the Higgs decay, and $b_3$, $b_4$ are the decay products of the other. We then demand that the corresponding jets reconstruct the two Higgs bosons within the on-shell window ($90~\text{GeV},130~\text{GeV}$). 

With the selected signal events, we construct a $\chi^2$ as described in Sec.\ \ref{subsub:statan}, again, scanning over $\kappa_V$ and $\kappa_{2V}$. The constraints for both our benchmark points are presented in Fig. \ref{fig:collider}, at integrated luminosities of $\mathcal{L}=1~\text{ab}^{-1}$ and $5~\text{ab}^{-1}$. Our analysis reproduces the CLIC projection~\cite{Roloff:2019crr},
\begin{equation}
  \kappa_{2V} \in [0.97,1.05] ,~\text{expected $95\%$ CL., 3 TeV, 5 ab${}^{-1}$.}
\end{equation}
However, the potential to exploit beam polarisations in \cite{Roloff:2019crr} (which we do not, here in our comparison) indicates that our sensitivity estimates are conservative.

\begin{figure*}[!t]
\centering
\includegraphics[width=0.48\textwidth]{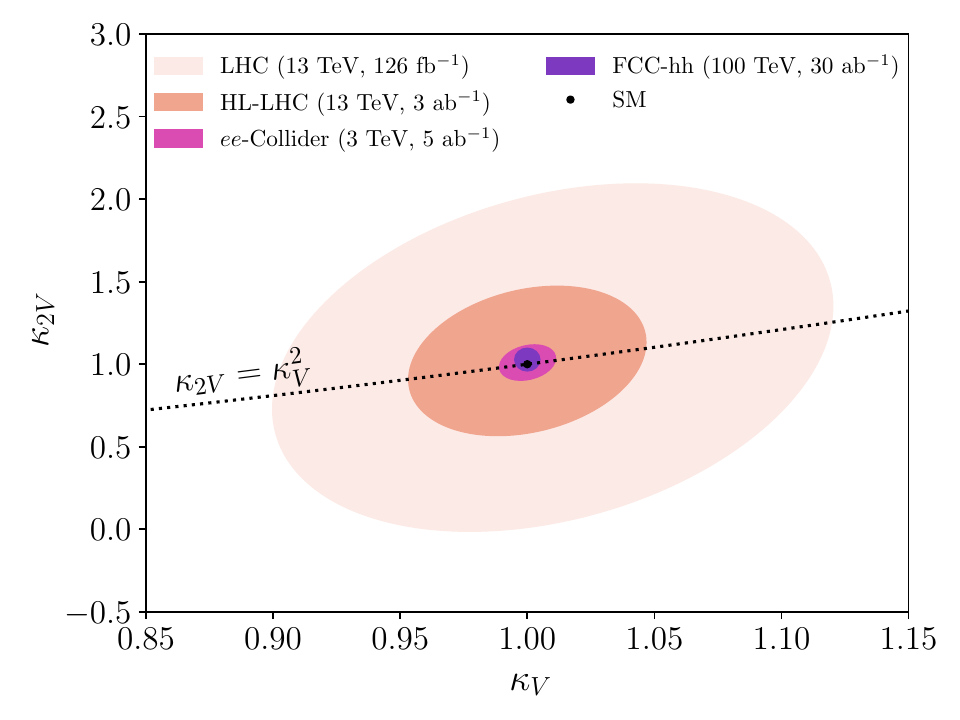}
\caption{\label{fig:newcollider} The 95\% CL constraints on $\kappa_V$ and $\kappa_{2V}$ parameter space resulting from the combination of WBF di-Higgs and single Higgs production for various collider options. Single Higgs constraints for future colliders are taken from the $\kappa_W$ bounds in \cite{deBlas:2019rxi}.}
\end{figure*}

In the following sections, we will treat the $\kappa_V$ single Higgs constraints independently from the $\kappa_{2V},\kappa_{\lambda}$ constraints for presentation purposes (in our discussion below, these limits correspond to qualitatively different phenomenological parameters). Of course, the individual rectangular regions indicated in Fig.~\ref{fig:collider} are simultaneously constrained by these (correlated) data sets. To gauge how the rectangular regions map onto the elliptical constraints in a combination of this information for the most sensitive environments, we show a combination of single Higgs and weak boson fusion $hh$ production in Fig.~\ref{fig:newcollider}. This demonstrates that in particular at the very sensitive future collider environments that enable tight constraints on $\kappa_{2V}$, this parameter will remain less constrained compared to $\kappa_V$.

\subsection*{Remarks on $\kappa_\lambda$}
Weak boson fusion processes at hadron colliders play a less relevant role in constraining Higgs self-interactions, predominantly the trilinear Higgs coupling,
as Higgs pair production from gluon fusion $gg \to hh$ is more abundant. Representative extrapolations to the HL-LHC phase~\cite{Cepeda:2019klc} indicate that 
this coupling could be constrained at 50\% around its SM expectation. The environment of an FCC-hh at 100 TeV will increase the sensitivity to 3-5\%~\cite{Contino:2016spe}.

Higgs self-coupling measurements at lepton colliders are dominated by $Z$-boson associated Higgs pair production for low centre-of-mass energies and WBF at large energies, \emph{e.g.}, at CLIC. The latter, maximising the WBF potential, has a projected sensitivity of about $\kappa_{\lambda}=1^{+0.25}_{-0.15}$ with some level of degeneracy between $\kappa_\lambda$ and $\kappa_{2V}$~\cite{Roloff:2019crr}. As for hadron colliders the constraint on $\kappa_{2V}$ is stronger compared to $\kappa_\lambda$ which highlights the need for constraining $\kappa_\lambda$ at hadron colliders such as the LHC. The sensitivity at CLIC compares to an estimated ILC sensitivity (in the 250~GeV+500~GeV combined phase) of $\kappa_\lambda = 1\pm 0.25$~\cite{ILC:2019gyn}. 

The main focus of our work is a discussion of $\kappa_{2V},\kappa_V$ in WBF as this channel is primarily sensitive to these parameters due to \cref{eq:highenergy}. We will include, however, details on $\kappa_\lambda$ and its relevance in comparison to $\kappa_{2V},\kappa_V$ in the discussion of the next \cref{sec:theory}. $\kappa_\lambda$ is known to be relatively efficiently constrainable in gluon fusion Higgs pair production through sensitivity in the threshold region, whose inclusive cross section~\cite{Baur:2002rb,Dolan:2012rv} is an order of magnitude larger than WBF. Our discussion of $\kappa_\lambda$ should therefore be understood as a potential additional constraint when considering $\kappa_{2V}$ measurements as performed by the experiments.

\section{$\kappa_V$, $\kappa_{2V}$ and $\kappa_\lambda$ in BSM models}
\label{sec:theory}

In this Section, we consider how different BSM models populate the space of $\kappa_V$, $\kappa_{2V}$, and $\kappa_\lambda$. We pay particular attention to those models that predict order-of-magnitude larger deviations in either $\kappa_{2V}$ or $\kappa_\lambda$ relative to $\kappa_V$; in such cases future measurements of either $\kappa_{2V}$ or $\kappa_\lambda$ are likely to provide significant discriminating power. As discussed recently in \cite{Alonso:2021rac}, and as we elucidate below, a large deviation in $\kappa_{2V}$ relative to $\kappa_{V}$ will require non-decoupling TeV scale new physics.

\subsection{Extended scalar sectors, tree level\label{subsec:treelevel}}

Consider a generic extended scalar sector, built out of a set $M$ of electroweak multiplets $\Phi_A$. Let $\Phi_A$ have an $SU(2)$ irreducible representation (irrep) of dimension $d_A$, and a hypercharge $Y_A$, with renormalisable lagrangian
\begin{equation}
  \lag = \sum_{A \in M} \abs{D \Phi_A}^2 - V(\Phi) \, .
  \label{eq:treeLevelLag}
\end{equation}
We define the charged current part of the covariant derivative to be
\begin{align}
  D_\mu \Phi_A =&  \left(\partial_\mu - \frac{1}{\sqrt{2}} i g_W ( W^+_\mu T_A^+ + W^-_\mu T_A^-) - \ldots \right) \Phi_A \, ,
  \label{eq:covDer}
\end{align}
where the components of the generators in the $SU(2)$ irrep of dimension $d_A$ are given by
\begin{equation}
  [T^\pm_A]_{\alpha\beta} = \begin{cases}
     \sqrt{ \left(\frac{d_A+1}{2} \right) (\alpha+\beta-1) - \alpha \beta} & \text{if } \alpha \pm 1=\beta \\
     0 & \text{otherwise}
  \end{cases} \, ,
  \label{eq:generatorNorm}
\end{equation}
with the $SU(2)$ indices $\alpha,\beta$ running between $1$ (labelling the component with maximum third component of isospin, $T^3$) and $d_A$ (labelling the component with minimum $T^3$).

The $\left[\left(\frac{d_A+1}{2}\right)+Y_A\right]$-th component is electrically neutral, and after electroweak symmetry breaking can be decomposed into vevs, $v_R$ and $v_I$, and vev-free fields $h_R$ and $h_I$
\begin{equation}
  [\Phi_A]_{\left(\frac{d_A+1}{2}\right)+Y_A} = \frac{1}{\sqrt{2}} (v_R + h_R) +  \frac{i}{\sqrt{2}} (v_I + h_I) \, .
  \label{eq:neutralComponent}
\end{equation}
These are separated into real ($R$) and imaginary ($I$) parts. The imaginary components are absent from real irreps, as well as, without loss of generality, from one complex irrep due to our freedom to gauge it away. Assuming the scalar sector contains a hypercharge-$\frac12$ doublet, we gauge away its imaginary component.

Substituting \cref{eq:covDer,eq:neutralComponent} into \cref{eq:treeLevelLag} we see that in the broken phase the charged current interactions of the neutral Higgses are governed by
\begin{equation}
  \lag = \sum_i \frac12 (\partial h_i)^2 -V(v,h) + \frac14 g_W^2 W^+ W^-  \left[ C_{ij} v_i v_j + 2 C_{ij} v_i h_j + C_{ij} h_i h_j  \right]  + \dots\, ,
  \label{eq:treeLevelLagBroken}
\end{equation}
where $i,j$ index the neutral Higgses, and repeated indices are summed over, and the matrix $C$ is diagonal with entries
\begin{equation}
  C_{ij} = \delta_{ij} \left( \frac12 (d_A^2-1) - 2 Y_A^2 \right) \, . \label{eq:CmatDef}
\end{equation}
$d_A,Y_A$ are the dimension and hypercharge of the multiplet from which the $i$th component came: real and imaginary components of the same multiplet have the same entry. $C_{ij}$ is normalised to be the identity matrix in an $n$-Higgs doublet model.

The 125 GeV Higgs, $h$, is defined as a unit direction in the space of neutral Higgs components:
\begin{equation}
  h_i = h \, \hat{n}_i\, , \text{ where } \hat{n}_i \hat{n}_i = 1 \, .
\end{equation}
We substitute the above into \cref{eq:treeLevelLagBroken} and compare with \cref{eq:kappaVand2Vdef}. Identifying the total vev as
\begin{equation}
  v \equiv \left(C_{ij} v_i v_j\right)^\frac12 = 246 \, \mathrm{GeV} \, ,
\end{equation}
we obtain the coupling modifiers
\begin{equation}
\begin{split}
  \kappa_V &= \frac{C_{ij} v_i \hat{n}_j}{\left(C_{ij} v_i v_j\right)^\frac12} \, , \\
  \kappa_{2V} &= C_{ij} \hat{n}_i \hat{n}_j \, .
\label{eqn:bsm_ren}
\end{split}
\end{equation}
Note that we have assumed the electroweak multiplets form complete custodial irreps, so that $\kappa_V=\kappa_W=\kappa_Z$ and $\kappa_{2V}=\kappa_{2W}=\kappa_{2Z}$.

$\kappa_V$ and $\kappa_{2V}$ are therefore correlated in particular extended scalar sectors, \emph{i.e.} for particular $C_{ij}$. We give the examples of the singlet, second Higgs doublet, and Georgi-Machacek model in \cref{tab:kappas} and \cref{app:specificScalars}. In \cref{fig:kv2v}, we show the lines and regions that these models can populate in the $\kappa_V$-$\kappa_{2V}$ plane. Notably, for \emph{any} tree-level model
\begin{equation}
  v^2 \left( \kappa_{2V} - \kappa_V^2\right) = C_{ij} \hat{n}_i \hat{n}_j \, C_{kl} v_k v_l - \left( C_{ij} v_i \hat{n}_j \right)^2 \geq 0 \, ,
\end{equation}
which follows from the Cauchy-Schwartz inequality when $C$ is a positive definite matrix. It is only zero in the alignment limit, when $\hat{n}_i \propto v_i$, or in the case of mixing with singlets, when $C$ is only positive semidefinite.

Thus, to obtain a large deviation in $\kappa_{2V}$ but not $\kappa_V$ in an extended scalar sector, we require electroweak triplets or higher representations, and a departure from the alignment limit (\emph{i.e.}, a departure from the parabola $\kappa_{2V} = \kappa_V^2$), see also~\cite{Gunion:1989ci}. This in turn implies significant mixing of components of the Higgs doublet with other states that cannot be made arbitrarily heavy. This is likely to cause some tension with direct searches. For instance, a significant triplet component to electroweak symmetry breaking introduces resonant tell-tale same-sign $WW$ WBF production~\cite{Englert:2013wga} which drives constraints on the triplet nature of the observed Higgs boson~\cite{CMS:2021wlt} (see also~\cite{Ismail:2020zoz}). 
Finally, we remark here that, in the decoupling limit, both $\kappa_V$ and $\kappa_{2V}$ approach $1$, and in principle do so from different directions in the $\kappa_V$-$\kappa_{2V}$ plane, depending on the model --- compare, for instance, the singlet and 2HDM trajectories in \cref{fig:kv2v}. However, in the decoupling limit, the deviations from the Standard Model in $WW\to hh$ due to $\kappa_V,\kappa_{2V}$ are comparable to the short-distance contributions from heavy Higgs exchange. These latter exchange contributions are calculated for the singlet and 2HDM model in \cite{Egana-Ugrinovic:2015vgy}, where this effect is discussed in detail. The effect of both $\kappa_{2V}$ and the short distance heavy Higgs exchange can be combined into a $\kappa_{2V}^\text{eff}$, which satisfies
\begin{equation}
  1 - \kappa_V^2 = \kappa_V^2 - \kappa_{2V}^\text{eff} \, ,
\end{equation}
in the decoupling limit. This corresponds to the pattern predicted by the SMEFT at dimension 6 (see, \emph{e.g.}, \cite{Alonso:2021rac}).

Turning now to $\kappa_\lambda$, this is generally a free parameter for renormalisable potentials $V(\Phi)$ containing cubic interactions among the electroweak irreps. However, if the cubic interactions are absent (as often happens accidentally due to the charges of the multiplets, or is imposed by certain $\mathbb{Z}_2$ symmetries), we can begin to understand the range of $\kappa_\lambda$ close to the alignment limit. The potential among the neutral components before electroweak symmetry breaking will have the generic form
\begin{equation}
  V= \frac12 \mu^2_{ij} r_i r_j + \frac14 \lambda_{ijkl} r_i r_j r_k r_l \, ,
\end{equation}
where the tensors $\mu^2$ and $\lambda$ are necessarily symmetric in their indices, and $r_i=0$ corresponds to the electroweak symmetry preserving vacuum. After electroweak symmetry breaking, substituting $r_i = v_i + h_i$ leads to
\begin{equation}
  V= \text{const.} + \frac12 \mu^2_{ij} h_i h_j + \frac32 \lambda_{ijkl} h_i h_j v_k v_l + \lambda_{ijkl} h_i h_j h_k v_l + \mathrm{O}(h^4)
  \label{eq:noCubicPot}
\end{equation}
where we've used the vev condition
\begin{equation}
  \mu^2_{ij} v_j + \lambda_{ijkl} v_j v_k v_l = 0 \, .
  \label{eq:vevcond}
\end{equation}
From comparison with \cref{eq:kappaVand2Vdef}, $\kappa_\lambda$ takes the form
\begin{equation}
  \kappa_\lambda = \left(C_{ij} v_i v_j \right)^\frac12 \frac{2\lambda_{ijkl} \hat{n}_i \hat{n}_j \hat{n}_k v_l}{ m_h^2}
  \label{eq:klambda}
\end{equation}

Close to the alignment limit, \cref{eq:klambda} can be expressed in terms of mass parameters and the degree of alignment. We work, without loss of generality, in the mass basis where the mass matrix of \cref{eq:noCubicPot} satisfies
\begin{equation}
\begin{split}
  \mu^2_{11} + 3 \lambda_{11kl} v_k v_l =& m_h^2 \\
  \mu^2_{1a} + 3 \lambda_{1akl} v_k v_l =& 0  \\
  \mu^2_{ab} + 3 \lambda_{abkl} v_k v_l =& m_a^2 \delta_{ab} 
\label{eq:massEqs}
\end{split}
\end{equation} 
where $a,b=2,3,\ldots$ label heavy Higgs directions, each with mass $m_a^2$. In this basis, $v_i$ satisfies
\begin{equation}
  v_i = v \left( 1- \frac12 \epsilon_a \epsilon_a , \, \epsilon_2, \, \epsilon_3, \ldots \right)^T + \mathrm{O}(\epsilon^3) \, ,
\end{equation}
for some small parameters $\epsilon_a$ describing the amount of the vev in the heavy Higgs directions when close to the alignment limit. Summation is implied over repeated $a$ indices.

Expanding $\mu^2_{11},\mu^2_{1a},\mu^2_{ab}$ and $\lambda_{111l} v_l,\lambda_{11al} v_l,\lambda_{1abl} v_l$ order-by-order in $\epsilon_a$, \cref{eq:vevcond,eq:massEqs} can be expanded to solve for
\begin{equation}
  \kappa_\lambda = \frac{2v \lambda_{111l} v_l }{m_h^2} = 1 + \frac12 \epsilon_a \epsilon_a - 2  \epsilon_a \epsilon_b \frac{\left( \lim_{\epsilon_a \to 0} \mu^2_{ab} \right)}{m_h^2} + \mathrm{O}(\epsilon^3) \, .
\end{equation}

In the decoupling limit, where $m_a^2$ are large, the model is necessarily aligned and $\epsilon_a \sim \mathrm{O}\left(\frac{1}{m_a^2}\right)$ are correspondingly small. The quartic couplings of the model are limited in size by perturbative unitarity, and so from \cref{eq:massEqs} $\mu^2_{ab}$ approaches $m_a^2 \delta_{ab}$. In this case
\begin{equation}
  \kappa_\lambda \approx 1  - 2 \sum_a \epsilon_a^2 \left( \frac{m_a^2}{m_h^2} - \frac14 \right) \, ,
\end{equation}
and the deviation in $\kappa_\lambda$ enjoys a parametric enhancement of $\frac{m_a^2}{m_h^2}$ relative to the deviations in $\kappa^2_{V}$ and $\kappa_{2V}$ which are both $\mathrm{O}(C \times \epsilon_a^2)$. We note that this enhancement does not necessarily happen in the case of \emph{alignment without decoupling}.

\subsection{Extended scalar sectors, loop level\label{subsec:looplevel}}

We turn to the case of an arbitrary electroweak scalar multiplet, $\Phi_A$, with a $\mathbb{Z}_2$ symmetry that prevents it from acquiring a vev, and a cross quartic interaction $\lambda$ with the SM Higgs doublet $\Phi$:
\begin{equation}
\lag  = \abs{D \Phi_A}^2 - m^2_\varphi \abs{\Phi_A}^2 - 2 \lambda \abs{\Phi_A}^2 \left(\Phi^\dagger \Phi - {v^2\over2 } \right) \, .
\label{eq:loopLevelLag}
\end{equation}
If the scalar is sufficiently heavy, $2 m_\varphi>m_h$, its leading order effects are at one-loop level. 
When augmented with a small amount of $\mathbb{Z}_2$ splitting to allow charged particles to decay, \cref{eq:loopLevelLag} presents a minimal class of models that includes multiple viable candidates for BSM particles of mass $m^2_\varphi \lesssim 1 \, \mathrm{TeV}$ \cite{Banta:2021dek}. If the cross quartic is large, such that $m^2_\varphi \sim \lambda v^2$, sizeable effects in the electroweak phase transition are expected \cite{Banta:2022rwg}.

In the simplest case of a $\mathbb{Z}_2$ symmetric singlet, $\sqrt{2} \Phi_A \sim \sqrt{2} \Phi_A^\dagger \sim S$ and \cref{eq:loopLevelLag} reads 
\begin{equation}
  {\cal{L}} = {\cal{L}}_\text{SM} + {1\over 2} (\partial_\mu S)^2 - {m_S^2\over 2} S^2 - \lambda S^2 \left(\Phi^\dagger \Phi -{v^2\over 2} \right)\,. \label{eq:loopSingletLag}
\end{equation}
In this scenario the $WW\to hh$ subamplitudes (as any amplitude) can be renormalised. We adopt the on-shell scheme as described in Ref.~\cite{Ross:1973fp}, which is the common scheme used for electroweak corrections (we are treating tadpoles as parameters as in Ref.~\cite{Denner:2018opp}).\footnote{Specifically, we choose $\{m_W, m_Z, \alpha\}$ as input parameters for the electroweak sector. The Weinberg angle is then a derived quantity~\cite{Sirlin:1980nh}.} The effects of the singlet, to order $\mathrm{O}(\lambda^3)$, can be accounted for by substituting the following finite expressions into~\cref{fig:diags}
\begin{equation}
\label{eq:singlet}
\begin{split}
\kappa_{2V} -1=\;&  - {\lambda^2 v^2\over 8\pi^2} \text{Re}\left[ B_0'(m_h^2,m_S^2,m_S^2) \right] \simeq \kappa_V^2-1\,,\\
\kappa_{\lambda} -1 =\;& - {\lambda^3 v^4 \over 6\pi^2 m_h^2} C_0 (M^2_{hh},m^2_h,m^2_S) \\ &- {\lambda^2 v^2\over 8\pi^2} {1\over M_{hh}^2 - m_h^2} 
\left( B_0(M_{hh}^2,m_S^2,m_S^2) - \text{Re}[B_0(m_h^2,m_S^2,m_S^2)] \right) \\
& - {\lambda^2 v^2 \over 48\pi^2 m_h^2} \big(
4 B_0(m_h^2,m_S^2,m_S^2) + 2 B_0(M_{hh}^2,m_S^2,m_S^2) \\ & \hspace{3cm} - 
       \text{Re}[ 6 B_0(m_h^2,m_S^2,m_S^2) - 3m_h^2 B'_0(m_h^2,m_S^2,m_S^2)] \big)\,.
\end{split}
\end{equation}
Here $B_0$, $B'_0$ are the Passarino-Veltman~\cite{Passarino:1978jh} two-point function and its derivative, respectively, and $C_0$ is the 3-point function. $\kappa_\lambda$ becomes an $M_{hh}^2$-dependent form factor; this momentum dependence is particularly useful for light scalar masses $m_S$ which do not admit a reliable EFT description and can modify the phenomenology at threshold.

\begin{figure}[!t]
\centering
\includegraphics[width=0.7\linewidth]{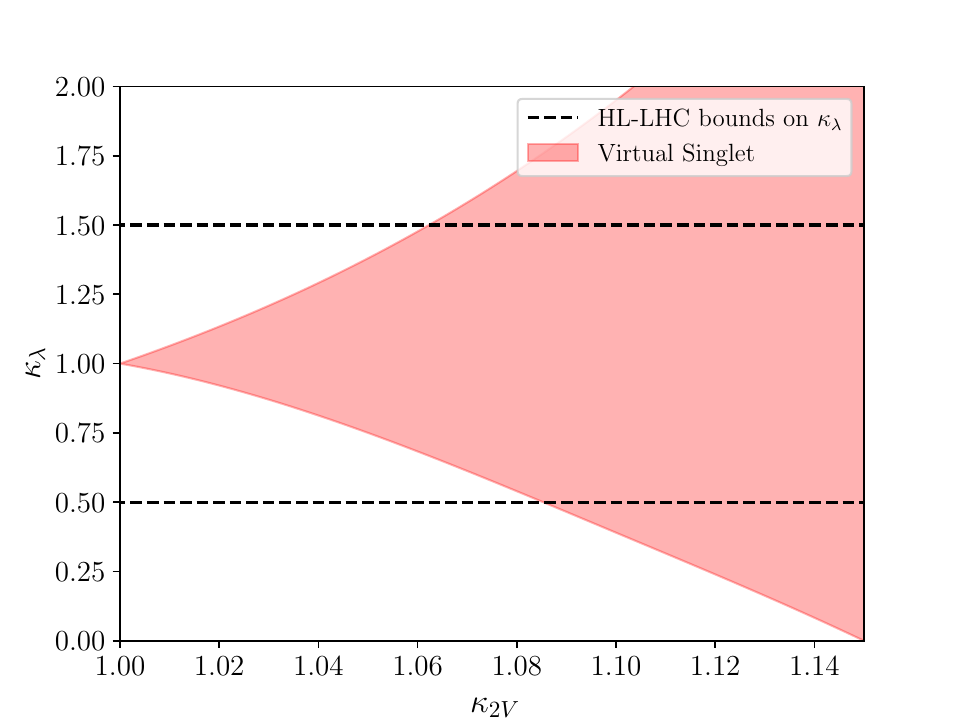}
\caption{\label{fig:singlet}$\kappa_\lambda$-$\kappa_{2V}$ plot for the aforementioned singlet mixing scenario for a representative value of $M_{hh}=300~\text{GeV}\gtrsim 2m_h$ which provides the region sensitive to $\kappa_\lambda$ investigations in Higgs pair production from gluon fusion. The HL-LHC projected sensitivity bounds on $\kappa_\lambda$ is represented by the dashed black lines~\cite{Cepeda:2019klc}. 
The range on the $\kappa_{2V}$ axis comes from HL-LHC bounds on $\kappa_V$, assuming $\kappa_{2V} = \kappa_V^2$ as in this singlet model. We vary $m_S\in [90,400]~\text{GeV}$ and $\lambda \in [-2,2]$ to obtain the contour.}
\end{figure}

Note that, for the loop-level singlet, the correction to $\kappa_V$ and $\kappa_{2V}$ is purely through the wavefunction renormalisation of the Higgs, and therefore follows a characteristic $\kappa_V^2 \simeq \kappa_{2V}$ pattern, the corrections to both scaling as $\lambda^2$. $\kappa_\lambda$, on the other hand, receives a $\lambda^3$ contribution from a 1PI singlet loop. When $\lambda$ is large, this generically means the $\kappa_\lambda$ sensitivity to this model at HL-LHC is greater than that of $\kappa_V,\kappa_{2V}$, as illustrated in \cref{fig:singlet}.

In principle, a $\Phi_A$ with non-trivial electroweak charges can contribute 1PI corrections to $hWW$ and $hhWW$ vertices; in practice, however, these states' corrections to the $\kappa$ parameters are parametrically similar to the singlet case. To see this, take $\Phi_A$ to be a second (inert) Higgs doublet as a representative example, and work to order $\mathrm{O}(m_\varphi^{-2})$, assuming the extra states are sufficiently heavy for their effects in $WW\to hh$ to be well approximated by constant $\kappa$s. Performing the same calculation as before in the on-shell scheme we obtain, for the off-shell SM-like 1 particle irreducible vertex functions $\Gamma$,
      \begin{align}
        \label{eq:loopDoubletKappas}
        { \Gamma^{WWh} \over \Gamma^{WWh}_\text{LO}} \bigg|_{\text{SM}}&=\bar \kappa_{W}=  1 - {g_W^2\over 1920\pi^2  m_\varphi^2} \left(2 {m_W^2}  + m_Z^2  \right) - {\lambda^2 v^2 \over 24\pi^2 m_\varphi^2} \, , \nonumber \\
        { \Gamma^{WWh} \over \Gamma^{WWh}_\text{LO}} \bigg|_{\text{SM}}&=\bar \kappa_{Z} = \bar \kappa_{W}   + {g_Y^2 \over 960 \pi^2 m_\varphi^2 } (m_W^2 -m_Z^2) \, , \nonumber\\
        { \Gamma^{WWhh} \over \Gamma^{WWhh}_\text{LO}} \bigg|_{\text{SM}}&=\bar \kappa_{2W}= 1  - {g_W^2\over 480\pi^2 m_{\varphi}^2}  {c_W^2\over s_W^2}  \left( {m_W^2}  + m_Z^2  \right)  - {\lambda^2 v^2 \over 12\pi^2 m_\varphi^2} \, , \\
        { \Gamma^{ZZhh} \over \Gamma^{ZZhh}_\text{LO}} \bigg|_{\text{SM}}&=\bar \kappa_{2Z}=\bar \kappa_{2W}  + {e^4 v^2 \over 1920\pi^2 m_\varphi^2} {g_Y^2 \over g_W^2} \, ,  \nonumber\\
        { \Gamma^{hhh} \over \Gamma^{hhh}_\text{LO}} \bigg|_{\text{SM}}&= \bar\kappa_\lambda  = 1  - {g_W^2 m_Z^2\over 1920\pi^2 m_{\varphi}^2}  - {\lambda^2 v^2\over 72\pi^2 m_\varphi^2} \left(7 + 2{M_{hh}^2\over m_h^2} \right) + {\lambda^3 v^2\over 3\pi^2 m_\varphi^2} {v^2\over m_h^2}\,.  \nonumber
      \end{align}
    We highlight the fact that these are no longer related to physical quantities by introducing the bar. 
    The physical couplings in the amplitudes are the product of both these corrections to the $\kappa$s, and the corrections to the electroweak couplings due to the non-trivial gauge quantum numbers of the new fields. In particular, for non-vanishing hypercharge, the new states are additional sources of custodial isospin violation as directly visible above. The gaugeless part of the corrections mirrors those of the singlet \cref{eq:singlet}. In particular, the $\mathrm{O}(\lambda^2)$ corrections to $\kappa_V$ and $\kappa_{2V}$ come exclusively from wavefunction normalisation of the Higgs boson.\footnote{Parametrically comparable 1PI contributions to $HWW$ and $HHWW$ arise from dimension-six operators $\sim |\Phi|^2 V_\mu V^\mu$. In cases where couplings are forbidden by gauge-invariance, \cref{eq:kappadef}, the dimension-6 contributions are conventionally included to the $\kappa$ definition as is the case for $\kappa_{\gamma},\kappa_g,\kappa_{Z\gamma}$, see \emph{e.g.}~\cite{Djouadi:2005gi,Carmi:2012in}.}
    
The above results clearly show that additional gauge interactions can sculpt the $\kappa_V,\kappa_{2V}$ parameter regions, however in a phenomenologically highly suppressed way compared to new additional inter-scalar interactions.
    Focussing on the gauge-independent part in practical $\kappa_{2V},\kappa_V,\kappa_\lambda$ analyses, the $\lambda$ contributions all scale with the number of real degrees of freedom, $D$. This gives approximate values for an arbitrary irrep of
      \begin{align}
        \kappa_{V} \simeq   \bar \kappa_{V}  & \approx 1 - D \, { \lambda^2 v^2 \over 96\pi^2 m_\varphi^2} \,, \nonumber \\
        \kappa_{2V} \simeq  \bar \kappa_{2V} &\approx   1  -  D \, {\lambda^2 v^2 \over 48\pi^2 m_\varphi^2} \,,\\
         \kappa_\lambda \simeq  \bar \kappa_\lambda & \approx 1  + D\, {\lambda^3 v^2\over 12\pi^2 m_\varphi^2} {v^2\over m_h^2}\,. \nonumber
       \end{align}
    where we have assumed $\lambda v^2 \gg m_h^2$ to drop the $\lambda^2$ piece in $\kappa_\lambda$, and focused on the HL-LHC measurement region $M_{hh}\gtrsim 2m_h$. The full momentum dependence for any such irrep can be restored from \cref{eq:singlet} by multiplying by the dimension $D$.

\subsection{From compositeness to dilaton mixing\label{subsec:compositeHiggs}}
The $\kappa$ values of a composite Higgs model depend on the details of the symmetry breaking. In the minimal composite Higgs model (MCHM), the components of the Higgs doublet chart the coset $SO(5)/SO(4)$ \cite{Contino:2003ve,Agashe:2004rs,Contino:2006qr}, whose relevant dynamics are readily constructed through the linear sigma model, see \cite{Alonso:2016btr}. The five components $\phi^1,\ldots,\phi^5$ have kinetic terms
\begin{equation}
  \lag = \frac12 \sum_{M=1}^4 \left((D \phi)^M \right)^2 + \frac12 (\partial \phi^5)^2 \, , \label{eq:mchmMetric}
\end{equation}
where the first four components have SM-like gauge couplings to the $W$ and $Z$. The components are restricted to the surface
\begin{equation}
  \sum_{M=1}^4 (\phi^M)^2 + (\phi^5)^2 = f^2
  \label{eq:MCHMConstraint}
\end{equation}
\newcommand{\shifth}{\mathfrak{h}}
which, in unitary gauge, can be parametrised by
\begin{equation}
  \left(\phi^1,\phi^2,\phi^3,\phi^4,\phi^5 \right) = \left(0,0,0,f \sin \frac{\shifth}{f} ,f \cos \frac{\shifth}{f} \right) \label{eq:mchmParam}
\end{equation}
where $\shifth$ is understood to be the Higgs coordinate shifted such that $\shifth=0$ in the absence of electroweak symmetry breaking. Substitution of \cref{eq:mchmParam} into \cref{eq:mchmMetric} yields the unitary gauge lagrangian
\begin{equation}
  \lag = \frac12 (\partial \shifth)^2 + {g_W^2 f^2 \over 4} \sin^2 \left( \shifth\over f \right) \left[  W^+ W^- + {1\over 2c_W^2} Z^2 \right]\,,
  \label{eq:MCHMWcouplings}
\end{equation}
and expanding about the vacuum $\shifth = \expv{\shifth}+h$ then gives
\begin{equation}
\begin{split}
  \kappa_{V} =& \sqrt{1-\xi}\,, \\
  \kappa_{2V} =& 1-2\xi\,,
\end{split}
\end{equation}
in terms of $\xi = \frac{v^2}{f^2}$. While this pattern will change for different cosets, it was shown in \cite{Alonso:2016oah} that for all custodial-symmetry-preserving cosets arising from the breaking of a compact group, it is guaranteed that
\begin{equation}
\begin{split}
  1 - \kappa_V^2 \geq& \, 0 \, , \\
  \kappa_V^2 - \kappa_{2V} \geq&\,  0 \, .
\end{split}
\end{equation}

By contrast, this can be violated for non-compact groups. A non-compact coset $SO(4,1)/SO(4)$ can be constructed via the linear sigma model lagrangian \cite{Alonso:2016btr}
\begin{equation}
  \lag = \frac12 \sum_{M=1}^4 \left((D \phi)^M \right)^2 - \frac12 (\partial \phi^5)^2 \, , \label{eq:hmchmMetric}
\end{equation}
restricted to the surface
\begin{equation}
  \sum_{M=1}^4 (\phi^M)^2 - (\phi^5)^2 = -f^2
  \label{eq:hyperbolicConstraint}
\end{equation}
which is parametrised in unitary gauge by
\begin{equation}
  \left(\phi^1,\phi^2,\phi^3,\phi^4,\phi^5 \right) = \left(0,0,0,f \sinh \frac{\shifth}{f} ,f \cosh \frac{\shifth}{f} \right) \, . \label{eq:hmchmParam}
\end{equation}
(Note that trigonometric functions in \cref{eq:mchmParam} become hyperbolic ones in \cref{eq:hmchmParam}, similar in spirit to the compactification of the Lorentz group.)
This yields
\begin{equation}
\begin{split}
  \kappa_{V} =& \sqrt{1+\xi}\,, \\
  \kappa_{2V} =& 1+2\xi\,,
\end{split}
\end{equation}
for the hyperbolic composite Higgs model.

In composite Higgs theories, the Higgs potential can be written schematically in MCHM5 (and MCHM5-like theories such as Ref.~\cite{Ferretti:2014qta}, where the 5 refers to the spurionic irrep of the top quark) as
\begin{equation}
\label{eq:compositeV}
f^{-4}V_\text{CH} \left( \frac{\shifth}{f} \right) =  \alpha \cos\left({2\shifth\over f}\right) - \beta \sin^2\left({2\shifth\over f}\right) =  {m_H^2 \over 8 v^2 (1-\xi)} \left[ \sin^2\left({\shifth\over f}\right) - \xi \right]^2 + V_0 \, ,
\end{equation}
where $V_0$ is a constant, see also \cite{Contino:2010rs}. The coefficients $\alpha,\beta$ are related to two and four-point functions of the underlying strongly interacting theory~\cite{Golterman:2015zwa,DelDebbio:2017ini} responsible for partial compositeness, and they can be replaced as a function of $\xi$ and the Higgs mass. 
These coefficients can in principle be inferred from lattice computations (for recent progress see~\cite{Ayyar:2018glg,DelDebbio:2022qgu}) to uncover realistic UV completions. 
The Higgs trilinear coupling modifier is then given by
\begin{equation}
\kappa_\lambda = {1-2\xi\over \sqrt{1-\xi}}
\end{equation}
The expression for the hyperbolic composite Higgs model is again obtained from the replacement $\xi\to -\xi$ \cite{Alonso:2016btr}.

We note that the above expressions satisfy 
\begin{equation}
\kappa_V \kappa_\lambda = \kappa_{2V}\,.
\end{equation} 
Therefore \cref{fig:diags}(a) reproduces the behaviour of \cref{fig:diags}(b) at leading order in the MCHM5 scenario. This is due to the symmetry breaking potential being of the same functional form as the interaction of the Goldstone bosons.\footnote{
In MCHM4 the potential reads $f^{-4}V_\text{CH}\left( \frac{\shifth}{f} \right)=  \alpha \cos\left(\frac{\shifth}{f}\right) - \beta \sin^2\left(\frac{\shifth}{f}\right)$ leading to $\kappa_\lambda = \sqrt{1-\xi}$, see also \cite{Grober:2010yv}. In this scenario, which suffers from tension with electroweak precision constraints, we also have $\kappa_V \kappa_\lambda \neq \kappa_{2V}$.}

Ultimately, the value of $\kappa_\lambda$ depends on the (spurionic) representations of the explicit $SO(4)$ symmetry breaking in the model. Larger representations, leading to higher order Gegenbauer polynomials of $\sin\left(\frac{\shifth}{f}\right)$ in the potential, can break the above correlation between $\kappa_\lambda$ and $\kappa_V,\kappa_{2V}$, and generally lead to a parametric enhancement in the deviations of $\kappa_\lambda$ from the Standard Model \cite{Durieux:2021riy}.

\subsubsection{Deforming the MCHM with a dilaton}

\begin{figure}	
  \centering
  \includegraphics[width=0.65\linewidth]{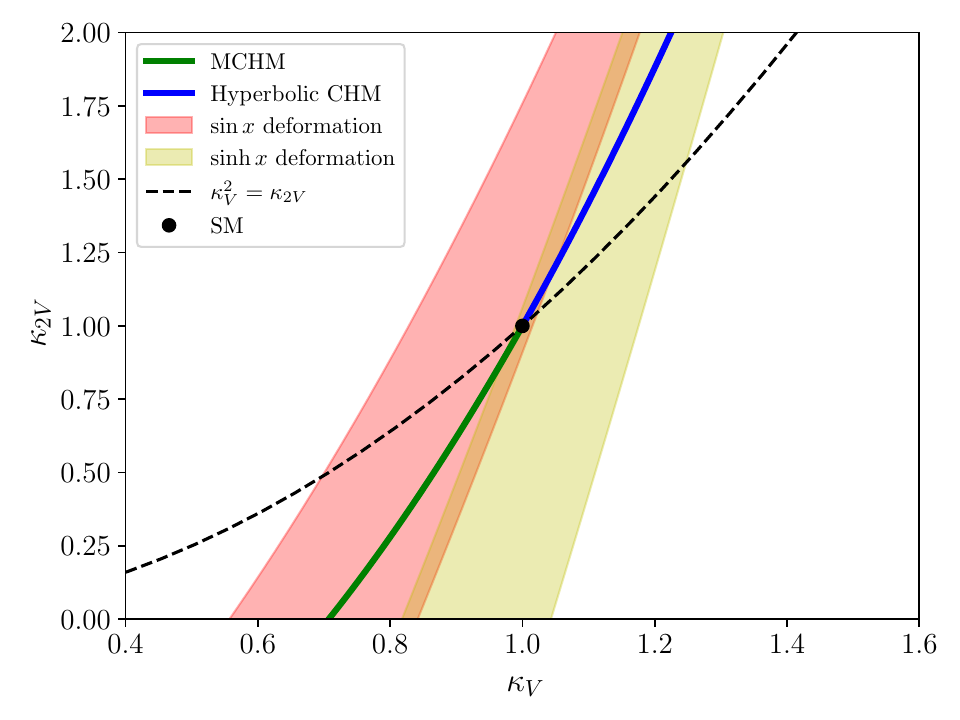}
  \caption{$\kappa_V$-$\kappa_{2V}$ predictions for the composite Higgs models and their deformation through Eq.~\eqref{eq:geometricDeformation} as described in the text. We scan $\kappa_V\in[0.5,1.5]$, $\xi =v^2/f^2 \in [0,1]$, and $f_t$ as a consistent solution of the $W$ mass given $\kappa_V,\xi$. This leaves $\kappa_{2V}$ determined as a function of these parameters. 
  \label{fig:composite}}
\end{figure}

The MCHM, together with its hyperbolic counterpart, define a line in the $\kappa_V$-$\kappa_{2V}$ plane (shown in green and blue in \cref{fig:composite}) corresponding to the maximally symmetric coset spaces of constant positive or negative curvature~\cite{alma9938889293902959}. Reference \cite{Alonso:2021rac} considered deformations away from this line that can arise from deformations of the coset space. Here, we begin similarly, and argue that a viable model for these deformations comes from mixing of the composite Higgs with a TeV scale dilaton.

Practically, the coset space can be deformed by replacing the constant radius $f$ in \cref{eq:MCHMConstraint,eq:hyperbolicConstraint} with a mildly $\phi^5$ dependent function
\begin{equation}
  f^2 \to f^2(\phi^5) \approx f^2_0 + f_0^2 {(\phi^5)^2\over f_t^2} \, ,
  \label{eq:geometricDeformation}
\end{equation}
for some $f_t^2 > f_0^2$. This promotes $\phi^5$ to a modulus that is common in higher dimensional theories of electroweak symmetry breaking.\footnote{The motivation 
of the scalar metric from which this theory derives also follows the discussion presented in \cite{Bekenstein:1992pj}. We have expanded to order $(\phi^5)^2$, assuming the linear term is forbidden by symmetry, so a linear deformation will not change our findings qualitatively.} We calculate the modifications $\kappa_V$, $\kappa_{2V}$ to linear order in $f_0^2/f_t^2$, after performing a field redefinition to canonically normalise the kinetic terms. (This field redefinition necessarily introduces momentum-dependent self-interactions of the would-be Higgs boson, to which we return below.) Scanning over $f_0^2/f_t^2$, we see from \cref{fig:composite} that such manifolds deviate from the line of uniform curvature in the $\kappa_V$-$\kappa_{2V}$ plane.

The effective $\phi^5$ dependence of $f$ can be brought about through mixing with a dilaton direction, $\chi$, which we analyse here using the approach of \cite{Bruggisser:2022ofg}. $\chi$ measures the departure from conformal symmetry through its vacuum expectation value $\langle \chi\rangle$~\cite{Goldberger:2007zk}. In a UV completion where the composite Higgs components are mesonic states arising from a confining gauge group $SU(N_c)$, $\expv{\chi} \sim f$ if the dilaton is another mesonic state, $\expv{\chi} \sim f \sqrt{N_c}$ if the dilaton is a glueball like state \cite{Bruggisser:2022ofg}. Here $\expv{\chi}$ is a free parameter, which we require to be $\mathrm{O}(\mathrm{TeV})$ in order to get sizeable mixing effects with the composite Higgs states.

Interaction terms with the composite Higgs can be constructed by multiplying operators by $\left(\chi/\langle \chi \rangle\right)^n$, where $n$ is the canonical mass dimension of the operator, in order to restore conformal symmetry. Thus
\begin{equation}
{\cal{L}}\supset {g_W^2 f^2 \over 4} \left({\chi\over\langle\chi\rangle}\right)^2 \sin^2 \left( \shifth \over f \right) \left[  W^+ W^- + {1\over 2c_W^2} Z^2 \right] - \left({\chi\over\langle\chi\rangle}\right)^4 V_\text{CH}\left(\shifth \over f \right)\,.
\end{equation}
in terms of the composite Higgs model operators \cref{eq:MCHMWcouplings,eq:compositeV}. Expanding this lagrangian around the minimum $\langle  \shifth \rangle,\langle \chi \rangle$ and including effects of Higgs-dilaton mixing through the usual isometry
\begin{equation}
\left(
\begin{matrix}
h\\
\chi - \expv{\chi}
\end{matrix}
\right)
=
\left(
\begin{matrix}
c_\phi & s_\phi\\
-s_\phi & c_\phi
\end{matrix}
\right)
\left(\begin{matrix}
h'\\
\chi'
\end{matrix}
\right)\,,
\end{equation}
we obtain, from the couplings of the mass eigenstate $h^\prime$,
\begin{equation}
\label{eq:klamd}
  \begin{split}
    \kappa_V &=c_\phi \sqrt{1-\xi}-s_\phi \sqrt{\zeta} \, , \\
    \kappa_{2V} &=(1-2\xi) c^2_\phi +\zeta s^2_\phi - 2\sqrt{\zeta (1-\xi)}s_{2\phi} \, , \\
    \kappa_\lambda &= c_\phi^3  {1-2\xi\over \sqrt{1-\xi}} - 4 c_\phi^2 s_\phi \sqrt{\zeta}-8 \frac{V_0}{m_h^2 v^2} s_\phi^3 \zeta^\frac32 \, ,
  \end{split}
\end{equation}
where $\zeta=v^2/\langle \chi\rangle^2$, and we have taken the MCHM5 potential for $V_\text{CH}$ from \cref{eq:compositeV}.

These couplings indeed cover, a priori, a wide area in the $\kappa_V$-$\kappa_{2V}$ plane. In particular, the area $\kappa_{2V}<1<\kappa_V$, which was not populated by the models considered in previous sections, can be reached
if the Higgs is considered mostly as a pseudo-dilaton. This area partially overlaps with the geometric deformations shown in \cref{fig:composite} (\cref{eq:geometricDeformation} can be thought of as approximating the leading order effects of dilaton mixing).
          
$\kappa_\lambda$ can also accommodate large deviations from $1$ in the case of significant Higgs-dilaton mixing. In addition, the dilaton $\chi$ has additional $\chi^3$ terms from explicit sources of conformal symmetry violation as described in~\cite{Rattazzi:2000hs,Goldberger:2007zk}. This means that trilinear interactions will receive momentum-dependent interactions as a consequence of the $a$-theorem~\cite{Komargodski:2011vj} leading to
\begin{equation}
{\cal{L}} \supset  {4\,\Delta a\over \langle \chi \rangle^3} \, (\partial \chi)^2\, \Box\, \chi + \dots
\end{equation}
where the ellipses denote higher order terms in the dilaton and $\Delta a = {\mathrm{O}}(\%)$ (see~\cite{Komargodski:2011vj,Dolan:2012ac}). 
This leads to an additional momentum-dependent modifications of $\kappa_\lambda$ of 
\begin{equation}
\label{eq:klamd2}
\Delta \kappa_\lambda =  s_\phi^3 {4\,\Delta a\over 3  }\,   \zeta \, \left[{M_{hh} \over \langle \chi\rangle}\right]^2 \,   \left[{M_{hh}  \over m_h} \right]^2 \,   \left(1-{4m_h^2\over M_{hh}^2} \right)  \,, 
\end{equation}
with $M_{hh}$ denoting the invariant di-Higgs mass. Modifications vanish close to the threshold, but can lead to a large modification of the invariant di-Higgs mass spectrum for considerable mixing.

The examples discussed so far exhaust the phenomenological possibilities in the $\kappa_V$-$\kappa_{2V}$ plane, and therefore provide a theoretical avenue to interpret the results of associated analyses. Of course, $\kappa_V$ constraints are also informed by single Higgs measurements and therefore there are
significant constraints on these scenarios from a range of experimental findings.
  Similarly, both momentum-dependent and momentum-independent modifications to $\kappa_\lambda$ are best constrained in measurements of gluon fusion Higgs pair production, given the larger 
  rate and the generic sensitivity of WBF to (multi-) gauge boson interactions.

\begin{table}[t!]
	\centering
	\renewcommand{\arraystretch}{1.7}
  \small{	\begin{tabular}{|c|c||c|c|}
		\hline 
		$\mathcal{O}_{HWW}$ 
		& $ -2a_{HWW}\;g_{W}^2 \frac{h}{v} \text{Tr}\Big[ W_{\mu \nu} W^{\mu \nu}\Big]$ 
		& $\mathcal{O}_{H {\mathcal{V}} {\mathcal{V}}}$ 
		& $-a_{H {\mathcal{V}} {\mathcal{V}}} \frac{m_h^2}{2}  \frac{h}{v} \text{Tr}\Big[ {\mathcal{V}}_{\mu} {\mathcal{V}}^{\mu}\Big]$ \\
		\hline
		$\mathcal{O}_{\square {\mathcal{V}} {\mathcal{V}} }$ 
		& $a_{\square {\mathcal{V}} {\mathcal{V}}}  \frac{\square h}{v} \text{Tr}\Big[ {\mathcal{V}}_{\mu} {\mathcal{V}}^{\mu}\Big]$ 
		& $\mathcal{O}_{H \square \square}$ 
		& $a_{H \square \square } \frac{h}{v} \frac{\square h \square h}{v^2}$ \\
		\hline
		$\mathcal{O}_{d2}$ 
		& $ i a_{d2} \; g_{W} \frac{\partial^{\nu} h}{v} \text{Tr}\Big[ W_{\mu \nu}{\mathcal{V}}^{\mu}\Big]$ 
		& $\mathcal{O}_{d d \square}$ 
		& $a_{d d \square }  \frac{\partial^\mu h \partial_\mu h }{v^3}$ \\
		\hline
		$\mathcal{O}_{\square \square}$ 
		& $a_{\square \square }  \frac{\square h \square h}{v^2}$ 
		& $\mathcal{O}_{H d d}$ 
		& $a_{H d d}  \frac{m_h^2}{v^2} \frac{h}{v} \partial^\mu h \partial_\mu h$ \\
		\hline
	\end{tabular}}
	\caption{HEFT operators $\mathcal{O}_{i}$ relevant for the RGE analysis, $a_{i}$ are the corresponding HEFT coefficients. ${\mathcal{V}}_{\mu}= (D_{\mu}U)U^{\dagger}$ and ${D}_{\mu}{\mathcal{V}}^{\mu}= \partial_{\mu}{\mathcal{V}}^{\mu} + i [g_{W}W_{\mu},{\mathcal{V}}^{\mu}] $. $W,B$ are the standard gauge field and field strengths. The non-linear sigma model parametrising the Goldstone fields is $U(\pi^{a}) = \exp\left({i \pi^{a} \tau^{a}/v}\right)$.} 
	\label{tab:operators}
\end{table}  

\subsection{Running of coupling modifiers}

Going beyond tree level, the correlations of $\kappa_V$, $\kappa_{2V}$, and $\kappa_\lambda$ become scale and scheme dependent.
In the Higgs effective field theory, the leading-order operators that modify $\kappa_V$, $\kappa_{2V}$, and $\kappa_\lambda$ (free, uncorrelated parameters in HEFT) not only run into each other, but also into higher derivative, next-to-leading-order operators that modify the measured $\kappa$s, as shown in \cref{tab:operators}. To encapsulate these effects, we define an effective $\kappa_V$ from the coefficient of $g^{\mu\nu}$ in the effective vertex for $hWW$ (as defined in \cite{Herrero:2022krh,Anisha:2022ctm})
\begin{equation}
  Z_h^\frac12 \hat\Gamma_{HW^-W^+}^{\mu\nu} \, ,
\end{equation}
with all legs on shell. The wavefunction normalisation of the Higgs is
\begin{equation}
  Z_h^{-1} = 1 + 4 \frac{m_h^2}{v^2} a_{\Box \Box}
\end{equation}
in terms of the operator coefficient $a_{\Box\Box}$ defined in \cref{tab:operators} \cite{Herrero:2022krh}. We obtain that
\begin{align}
  \kappa_V^\text{eff} =& \kappa_V + \frac{m_h^2}{v^2} \left( 2 a_{HWW} + a_{d2} + 2 a_{\Box \mathcal{V}\mathcal{V}} - a_{H\mathcal{V}\mathcal{V}} - 2 \kappa_V a_{\Box\Box}\right) -  \frac{m_W^2}{v^2} 4 a_{HWW}
\end{align}
Similarly, from the value of
\begin{equation}
  Z_h^\frac32 \hat\Gamma_{HHH} 
\end{equation}
with all legs on shell we define an effective coupling
\begin{equation}
  \kappa_\lambda^\text{eff} =  \kappa_\lambda + \frac{m_h^2}{v^2} \left(-2 a_{H\Box\Box} + a_{dd\Box}-a_{Hdd}-6 \kappa_\lambda a_{\Box\Box}\right) \, .
\end{equation}
These effective couplings serve the purpose of effectively field-redefining the redundant higher derivative operators generated by the running back into $\kappa_V$ and $\kappa_\lambda$ respectively. We do not define an analogous on-shell $\kappa_{2V}^\text{eff}$, as this would be a complicated function of components of the $hhWW$ effective vertex and components of diagrams containing, \emph{e.g.}, $hWW$ and $hhh$ vertices.

Using the results of \cite{Herrero:2021iqt,Herrero:2022krh} (see also~\cite{Delgado:2013hxa,Gavela:2014uta,Asiain:2021lch,Gomez-Ambrosio:2022why,Gomez-Ambrosio:2022qsi,Anisha:2022ctm}), the couplings run according to
  \begin{align}
    16 \pi^2 \frac{\dd}{\dd \log \mu^2} \kappa_V^\text{eff} =& \, \frac{m_h^2}{v^2} \left( \frac32 (\kappa_{2V}-\kappa_V^2)(\kappa_\lambda-\kappa_V) - \kappa_V(1-\kappa_V^2) \right)  \nonumber \\
    &+  \frac{m_W^2}{2 v^2} \left( 3\kappa_V(1-\kappa_V^2) + \frac{20}{3} \kappa_V (\kappa_{2V}-\kappa_V^2)\right)  \nonumber  \\
    &+  \frac{m_Z^2}{2 v^2}  3\kappa_V(1-\kappa_V^2)  \, ,\nonumber \\
    16 \pi^2 \frac{\dd}{\dd \log \mu^2} \left[ \kappa_{2V} - (\kappa_V^\text{eff})^2 \right] =& \, \frac{m_h^2}{2 v^2} \Big( 2 (\kappa_{2V}-\kappa_V^2)^2  + 4 \kappa_V^2 (1-\kappa_V^2) \nonumber  \\
    &\hspace{7ex} - (\kappa_{2V}-\kappa_V^2) (4\kappa_{2V}-16 \kappa_V^2+18 \kappa_\lambda \kappa_V-3 \kappa_4)\Big)\nonumber   \\
    &+ \frac{m_W^2}{ v^2} \left( 3 (\kappa_{2V}-\kappa_V^2)^2 + (3-\frac{20}{3}\kappa_V^2)(\kappa_{2V}-\kappa_V^2)\right) \nonumber \\
    & +  \frac{m_Z^2}{2 v^2}  3(\kappa_{2V}-\kappa_V^2)(1-\kappa_V^2)  \, , \nonumber \\
    16 \pi^2 \frac{\dd}{\dd \log \mu^2} \kappa_\lambda^\text{eff} =& \, \frac{m_h^2}{2 v^2} \left( 9 (\kappa_\lambda-\kappa_V) \kappa_\lambda^2 + (\kappa_\lambda+3 \kappa_V) (\kappa_{2V}-\kappa_V^2) + 6 \kappa_\lambda \kappa_4 - 6 \kappa_V^3  \right)  \nonumber \\
    & +  \frac{3 (m_W^2+m_Z^2)}{2 v^2} 3 \kappa_\lambda (1-\kappa_V^2)  \, . 
  \label{eq:RGs}
  \end{align}
$\kappa_4$ is the multiplicative modifier of the $h^4$ vertex, and is set to one in the following. We have chosen to present the running of $\kappa_{2V}$ in terms of that of the combination $\kappa_{2V} - (\kappa_V^\text{eff})^2$, which controls the energy growth of the $WW\to hh$ process.

The Standard Model is a fixed point of the running, as the RHSs of \cref{eq:RGs} vanish when $\kappa_V=\kappa_{2V}=\kappa_\lambda=1$.\footnote{Without defining $\kappa_\lambda^\text{eff}$ to take account of the effect of the higher derivative operators, $\kappa_\lambda$ would appear to run even at the Standard Model point.} Linearising the RGEs about this point by defining
\begin{equation}
\begin{split}
  \delta \kappa_V =& \kappa^\text{eff}_V -1 \, , \\
  \delta K_{2V} =& \kappa_{2V} -(\kappa_V^\text{eff})^2 \, ,\\
  \delta \kappa_\lambda =& \kappa_\lambda^\text{eff} -1 \, ,
\end{split}
\end{equation}
we find
\begin{equation}
  \label{eq:linearisedRGs}
  \begin{split}
    16 \pi^2 \frac{\dd}{\dd \log \mu^2} \delta \kappa_V =& \delta \kappa_V \left(2 \frac{m_h^2}{v^2} - \frac{3 (m_W^2+m_Z^2)}{2 v^2} \right) + \delta K_{2V} \left( \frac{3 m_W^2}{v^2} \right) \\
    16 \pi^2 \frac{\dd}{\dd \log \mu^2} \delta K_{2V} =& \delta \kappa_V \left(-4 \frac{m_h^2}{v^2}\right) + \delta K_{2V} \left( -\frac{3 m_h^2}{2v^2} - \frac{40 m_W^2}{3 v^2} \right) \\
    16 \pi^2 \frac{\dd}{\dd \log \mu^2} \delta \kappa_\lambda =& \delta \kappa_V \left(-\frac{27 m_h^2}{2 v^2} - \frac{9 (m_W^2+m_Z^2)}{2 v^2} \right) + \delta K_{2V} \left( \frac{2 m_h^2}{v^2} \right) + \delta \kappa_\lambda \left( \frac{15 m_h^2}{2 v^2} \right) 
  \end{split}
  \end{equation}
  up to $\mathrm{O}((\delta \kappa)^2)$ corrections. Note that, to expand the RHS of \cref{eq:RGs}, we have assumed we are running from a point where all higher order $a$ coefficients are zero.

We note from \cref{eq:linearisedRGs} that $\delta \kappa_\lambda$, which is experimentally the least constrained of the three parameters, self-renormalises significantly stronger than the other two. Also, the negative coefficient for the self-renormalisation of $\delta K_{2V}$ means that a positive $\delta K_{2V}$, as is the case for all renormalisable models, grows in the IR, away from the $\delta K_{2V}=0$ alignment limit.

\renewcommand{\arraystretch}{1.1} 
\begin{table}[!b]
  \centering
  \begin{footnotesize}
  \begin{tabular}{p{2.8cm} || p{4.cm} | p{4.4cm} | c}
    Model & $\kappa_V$ & $\kappa_{2V}$  & Ref. \\ \hline\hline
    Singlet & $\cos \alpha$ & $\cos^2 \alpha$ & \S\ref{subsec:singlet} \\ \hline
    2HDM & $\sin (\alpha-\beta)$ & $1$  & \S\ref{subsec:2HDM} \\ \hline    
    Georgi-Machacek & $\cos \alpha \cos \beta + 2 \sqrt{\frac23} \sin\alpha \sin\beta$ & $\cos^2 \alpha + \frac83 \sin^2\alpha$ & \S\ref{subsec:GM} \\ \hline 
    Tree-level scalar & $\displaystyle \frac{C_{ij} v_i \hat{n}_j}{\left(C_{ij} v_i v_j\right)^\frac12}$ & $C_{ij} \hat{n}_i \hat{n}_j$  & \S\ref{subsec:treelevel} \\ \hline
    Loop-level scalar& \multirow{2}{*}{$\displaystyle 1 - D \, { \lambda^2 v^2 \over 96\pi^2 m_\varphi^2}$} & \multirow{2}{*}{$\displaystyle 1  -  D \, {\lambda^2 v^2 \over 48\pi^2 m_\varphi^2}$}  & \multirow{2}{*}{\S\ref{subsec:looplevel}} \\ 
     (large $\lambda$) & & & \\ \hline
    \hline
    SMEFT & free & $\simeq 2\kappa^2_V-1$  &  \emph{e.g.}~\cite{Grzadkowski:2010es,Dedes:2017zog} \\   \hline
    MCHM & $\sqrt{1-\xi}$ & $1 - 2 \xi$ &  \S\ref{subsec:compositeHiggs} \\ \hline
    MCHM + Dilaton & $c_\phi \sqrt{1-\xi}-s_\phi \sqrt{\zeta}$ & $(1-2\xi) c^2_\phi +\zeta s^2_\phi - 2\sqrt{\zeta (1-\xi)}s_{2\phi}$ & \S\ref{subsec:compositeHiggs} \\  \hline
    HEFT & free & free & \emph{e.g.}~\cite{Herrero:2021iqt} \\ \hline
  \end{tabular}
  \caption{A collection of $\kappa_V,\kappa_{2V}$ values of the simplified models described in \cref{sec:theory} and \cref{app:specificScalars}.\label{tab:kappas}}
  \end{footnotesize}
\end{table}

\begin{figure}[!t]
\centering
\includegraphics[width=0.96\textwidth]{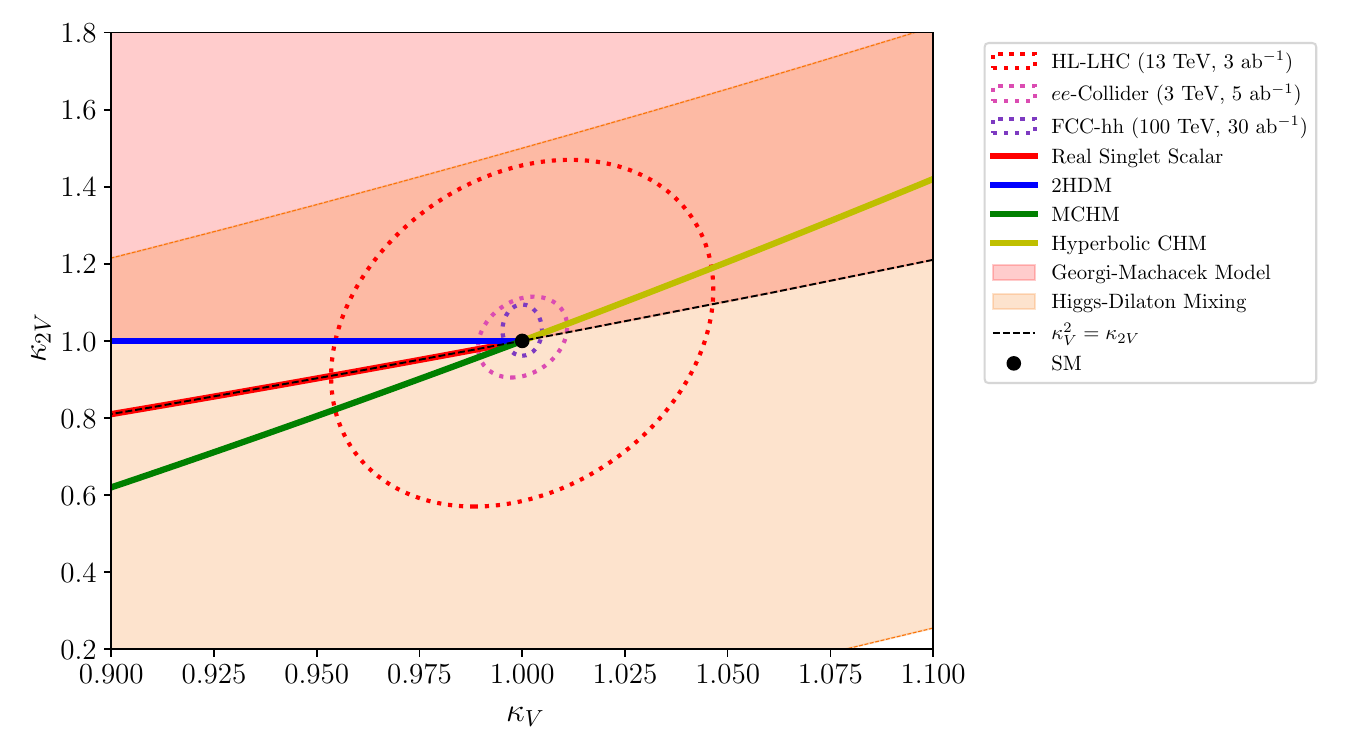}
\caption{$\kappa_V$-$\kappa_{2V}$ correlation for different 95\% CL collider sensitivity extrapolations, assuming $\kappa_\lambda=1$. We overlay the BSM model discussion of \cref{sec:theory} to 
highlight regions for which $\kappa_{2V}$ can provide information beyond $\kappa_V$. In particular, the fourth quadrant (where $\kappa_{2V}<1<\kappa_V$) is populated by scenarios
of large dilaton-Higgs mixing (we scan $-1\leq \kappa_{2V}\leq 2,|s_\theta|\leq 1$ with a physical solution of $\zeta>0$; again we assume $M_{hh}=300~\text{GeV}$ as before). Note that the dilaton includes effects beyond the deformation truncated in Eq.~\eqref{eq:geometricDeformation}. Hence, the covered area is comparably larger. The black dashed line represents $\kappa_V^2=\kappa_{2V}$ and all BSM renormalisable models lie in the region above this line, \emph{i.e.}, $\kappa_V^2\leq\kappa_{2V}$.
\label{fig:kv2v}}
\end{figure}

\section{Conclusions}
\label{sec:conc}
Analyses employing Feynman diagrams as templates are established approaches in the experimental community~\cite{CMS:2022nmn,ATLAS:2023qzf} in situations
where data is expected to be limited, even at the high-luminosity phase of the LHC or stretches of future collider runs. What can be learned from them?
To answer this question and provide an ab initio motivation for these analyses, we have performed such a templated analysis of future sensitivity in the WBF di-Higgs channel to the coupling modifiers $\kappa_{2V},\kappa_{V},\kappa_\lambda$, and we have scanned a range of BSM models to understand their pattern of effects in the same three coupling modifiers.

\cref{fig:kv2v} shows projected future constraints on the $\kappa_V$-$\kappa_{2V}$ plane from the WBF di-Higgs process. This translates into approximate $2 \sigma$ bounds on deviations in the single parameter $\kappa_{2V}$ of order $30\%$ for HL-LHC, and order $5\%$ for future colliders. These are to be contrasted with bounds on $\kappa_V$ from single Higgs measurements that are approximately an order of magnitude smaller.

On the theoretical side, deviations in $\kappa_{2V}$ and $\kappa_V$ are highly correlated in all models of heavy, decoupling new physics. This can be understood through their SMEFT parametrisation, where, due to the doublet nature of the Higgs field, $\kappa_{2V} =2 \kappa_V^2 - 1$ at dimension 6. In the broader HEFT parametrisation, $\kappa_{2V}$ and $\kappa_V$ are independent free parameters due to the custodial iso-singlet parameterisation of the Higgs boson. Any models of non-decoupling new physics where the deviation in $\kappa_{2V}$ is an order of magnitude larger than that in $\kappa_V$ would highly motivate WBF di-Higgs production as an indirect probe of new physics.

To this end, we surveyed many BSM models in \cref{sec:theory}, and summarise their patterns of effects in $\kappa_{2V}$ and $\kappa_V$ in \cref{tab:kappas}. In sum, to achieve an enhancement in $\kappa_{2V}$ within an extended scalar sector requires tree-level mixing with triplets or higher electroweak representations. This cannot be done at loop level, where extra scalars give a characteristic $\kappa_{2V} =\kappa_V^2$ pattern from wavefunction renormalisation of the Higgs. We note in passing, however, that we have identified a large class of scalar extensions of the SM for which a Feynman-templated analysis extends to one-loop BSM precision in the weak sector.

We find that all renormalisable extended scalar sectors satisfy $\kappa_{2V}-\kappa_V^2 \geq 0$, with this quantity growing in the IR due to running effects. As $\kappa_{2V}-\kappa_V^2$ also controls the energy growth of longitudinal $W$-Higgs scattering, it is possible that a sum rule type argument could yield further insight into this sign. $\kappa_{2V}-\kappa_V^2 < 0$ is achievable, however, in non-renormalisable models of the scalar sector. The archetype of such a model --- the minimal composite Higgs model --- follows the SMEFT pattern, as it smoothly decouples in the $f \to \infty$ limit. However, it is possible to obtain a large (and negative) deviation in $\kappa_{2V}$ in composite Higgs models which contain a TeV scale dilaton.

Our investigation therefore shows that the Feynman diagram-based explorations of the WBF channel add important value to the phenomenology programme, at the LHC and beyond. $\kappa_{2V}$ can significantly depart from the SM correlation with $\kappa_V$, and the loose constraints on $\kappa_{2V}$ probe parameter regions of BSM scenario that are not accessible by a precision study of $\kappa_V$ alone. This is assisted by considerable stability of the WBF cross sections from the point of view of QCD. However, measurable deviations in  $\kappa_{2V}$ come at the price of significant tree-level mixing at the TeV scale in either perturbative (extended scalar sectors containing higher electroweak irreps) or non-perturbative (composite Higgs with dilaton) scenarios. The size of tree-level mixing highlights the relevance of direct searches. To compare the power of direct searches, let us briefly consider the prospects at the LHC for the Georgi-Machacek and composite Higgs with dilaton models.

For the Georgi-Machacek model, tell-tale and experimentally clean signatures arise from the doubly charged, gauge-philic Higgs boson, whose interactions are sensitive to the triplets' contribution to the electroweak vacuum. Recent constraints~\cite{ATLAS:2023dbw} limit $|\sin\beta| \lesssim 0.2$ for masses below 1 TeV. A luminosity-based extrapolation should further decrease this by a factor of two for HL-LHC. Using the expressions in \cref{tab:kappas} with $|\sin\beta| \lesssim 0.2 (0.1)$ and $\kappa_V=1$ gives a possible range of $1 \leq \kappa_{2V} \lesssim 1.5(1.15)$ from current (future) LHC resonance searches, which is comparable to the $\kappa_{2V}$ constraints from di-Higgs measurements.

For the composite Higgs models, including those which mix with a dilaton, one expects other composite resonances to appear at a scale below $4 \pi f \gtrsim 5 \, \text{TeV}$, assuming $f \gtrsim 500 \, \text{GeV}$. Arguably, the most model independent predictions of the composite Higgs do not yet meet this threshold: the pair production of top partners is currently bounded up to masses of around $1.5 \, \text{TeV}$ \cite{ATLAS:2022hnn}, and the electroweak production and decay of a heavy vector triplet --- analogous to $\pi\pi \to \rho \to \pi\pi$ in QCD --- is currently bounded up to masses of around $4 \, \text{TeV}$ \cite{ATLAS:2020fry}. This latter bound is limited by the collision energy of the LHC, and is unlikely to improve significantly with increased luminosity. Therefore, using the expressions in \cref{tab:kappas} with $f,\expv{\chi} \gtrsim 500 \, \text{GeV}$ and $\kappa_V =1$ gives a possible range of $0.0 \lesssim \kappa_{2V} \lesssim 1.3$. We note that, if the dilaton has (model dependently) a coupling to gluons, a direct search for the dilaton in $gg \to \chi \to WW/ZZ$ could rule out a significant amount of this parameter space \cite{Bruggisser:2022ofg}.

Finally, although $\kappa_\lambda$ plays a subdominant phenomenological role in WBF $hh$ production in a hadron collider, in many of the scenarios that we have considered in this work, $\kappa_\lambda$ more susceptible to deviations from the SM compared to $\kappa_{2V},\kappa_V$. Examples include tree-level extended scalar sectors in the alignment with decoupling limit, loop-level scalars with sizeable cross-quartic interactions with the Higgs doublet, and composite Higgs models containing explicit symmetry breaking terms in higher spurionic irreps. As $\kappa_\lambda$ can be dominantly extracted from $gg\to hh$ analyses, correlations observed in the WBF mode will add further exclusion potential.

\acknowledgments

We thank Aidan Robson and Panagiotis Stylianou for helpful conversations.
C.E.\ is supported by the STFC under grant ST/T000945/1, the Leverhulme Trust under grant RPG-2021-031. C.E.\ and D.S.\ acknowledge support from the Institute for Particle Physics Phenomenology Associateship Scheme. W.N.\ is funded by a University of Glasgow College of Science and Engineering Scholarship.

\appendix
\section{$\kappa_V$ and $\kappa_{2V}$ in specific scalar models\label{app:specificScalars}}

Here we give the $\kappa_V$ and $\kappa_{2V}$ values at tree level in specific extended scalar sectors, using the master formulae \cref{eq:CmatDef,eqn:bsm_ren}.

\subsection{Higgs + singlet\label{subsec:singlet}}

  In unitary gauge, there is one neutral component of the Higgs doublet and one from the singlet, with $C$ matrix
  \begin{equation}
    C_{ij} = \begin{pmatrix}
      1 & 0 \\ 0 & 0
    \end{pmatrix} \, .
  \end{equation}
  Writing $v_i = \left(v_1,v_2\right)^T$ and $\hat{n}_i = \left(\cos \alpha,\sin \alpha\right)^T$, we have
  \begin{align}
    \kappa_V =& \cos \alpha \, , \\
    \kappa_{2V} =& \cos^2 \alpha \, ,
  \end{align}
  as expected.

  \subsection{2HDM\label{subsec:2HDM}}

  In unitary gauge there are two scalar neutral components, which both generically contain vevs and mix among each other, and a pseudoscalar neutral component which, when custodial symmetry is imposed, obtains no vev and does not mix with the other neutral components~\cite{Gunion:1989ci} (see the recent~\cite{Arco:2023sac} for the 2HDMs relation to HEFT). The $C$ matrix is the identity matrix,
  \begin{equation}
    C_{ij} = \begin{pmatrix}
      1 & 0 & 0 \\ 0 & 1 & 0 \\ 0 & 0 & 1
    \end{pmatrix} \, .
  \end{equation}
  Writing
  \begin{align}
    \hat{n}_i =& \left( \sin \alpha, \cos\alpha, 0 \right)^T \\
    v_i =& v \left( \cos \beta, \sin \beta, 0 \right)^T   
  \end{align}
  where the third entry is associated with the pseudoscalar component, we obtain
  \begin{align}
    \kappa_V =& \sin(\alpha-\beta) \, , \\
    \kappa_{2V} =& 1 \, .
  \end{align}

  \subsection{Georgi-Machacek\label{subsec:GM}}

  Ordering the neutral components respectively as that of the Higgs doublet, the two components of the complex $Y=1$ triplet and the one component of the real $Y=0$ triplet, the $C$ matrix is
  \begin{equation}
    C_{ij} = \begin{pmatrix}
      1 & 0 & 0 & 0 \\ 0 & 2 & 0 & 0 \\ 0 & 0 & 2 & 0 \\ 0 & 0 & 0 & 4
    \end{pmatrix}
  \end{equation}
  Custodial symmetry implies that $v_2=v_3=v_4$ among the triplet components, and also that the mass matrix is invariant under permutations among the 2, 3 and 4 indices, meaning $n_2=n_3=n_4$ in the light Higgs eigenvector \cite{Hartling:2014zca}. Thus we can write
  \begin{align}
    \hat{n}_i =& \left( \cos \alpha,\frac{1}{\sqrt{3}} \sin \alpha,\frac{1}{\sqrt{3}} \sin \alpha,\frac{1}{\sqrt{3}} \sin \alpha \right)^T \\
    v_i =& v \left( \cos \beta , \frac{1}{2 \sqrt{2}} \sin \beta ,\frac{1}{2 \sqrt{2}} \sin \beta ,\frac{1}{2 \sqrt{2}} \sin \beta \right)^T
  \end{align}
  which implies that
  \begin{align}
    \kappa_V =& \cos\beta \cos\alpha + 2 \sqrt{\frac23} \sin\beta \sin \alpha \, , \\
    \kappa_{2V} =& \cos^2 \alpha + \frac83 \sin^2 \alpha \, .
  \end{align}
  We have also reproduced this result using the mass matrices of \cite{Hartling:2014zca}, see also \cite{Alonso:2021rac}.

\bibliographystyle{JHEP}
\bibliography{references}

\providecommand{\href}[2]{#2}\begingroup\raggedright\begin{thebibliography}{100}

\bibitem{LHCHiggsCrossSectionWorkingGroup:2011wcg}
{\scshape LHC Higgs Cross Section Working Group} collaboration, S.~Dittmaier
  et~al., \emph{{Handbook of LHC Higgs Cross Sections: 1. Inclusive
  Observables}},  \href{https://arxiv.org/abs/1101.0593}{{\ttfamily
  1101.0593}}.

\bibitem{ATLAS:2023qzf}
{\scshape ATLAS} collaboration, G.~Aad et~al., \emph{{Search for nonresonant
  pair production of Higgs bosons in the $b\bar{b}b\bar{b}$ final state in $pp$
  collisions at $\sqrt{s}= 13$ TeV with the ATLAS detector}},
  \href{https://arxiv.org/abs/2301.03212}{{\ttfamily 2301.03212}}.

\bibitem{ATLAS:2015sxd}
{\scshape ATLAS} collaboration, G.~Aad et~al., \emph{{Searches for Higgs boson
  pair production in the $hh\to bb\tau\tau, \gamma\gamma WW^*, \gamma\gamma bb,
  bbbb$ channels with the ATLAS detector}},
  \href{http://dx.doi.org/10.1103/PhysRevD.92.092004}{\emph{Phys. Rev. D}
  {\bfseries 92} (2015) 092004},
  [\href{https://arxiv.org/abs/1509.04670}{{\ttfamily 1509.04670}}].

\bibitem{CMS:2017hea}
{\scshape CMS} collaboration, A.~M. Sirunyan et~al., \emph{{Search for Higgs
  boson pair production in events with two bottom quarks and two tau leptons in
  proton\textendash{}proton collisions at $\sqrt s$ =13TeV}},
  \href{http://dx.doi.org/10.1016/j.physletb.2018.01.001}{\emph{Phys. Lett. B}
  {\bfseries 778} (2018) 101--127},
  [\href{https://arxiv.org/abs/1707.02909}{{\ttfamily 1707.02909}}].

\bibitem{CMS:2017yfv}
{\scshape CMS} collaboration, A.~M. Sirunyan et~al., \emph{{Search for Higgs
  boson pair production in the $bb\tau\tau$ final state in proton-proton
  collisions at $\sqrt{s}$ = 8 TeV}},
  \href{http://dx.doi.org/10.1103/PhysRevD.96.072004}{\emph{Phys. Rev. D}
  {\bfseries 96} (2017) 072004},
  [\href{https://arxiv.org/abs/1707.00350}{{\ttfamily 1707.00350}}].

\bibitem{CMS:2020tkr}
{\scshape CMS} collaboration, A.~M. Sirunyan et~al., \emph{{Search for
  nonresonant Higgs boson pair production in final states with two bottom
  quarks and two photons in proton-proton collisions at $ \sqrt{s} $ = 13
  TeV}}, \href{http://dx.doi.org/10.1007/JHEP03(2021)257}{\emph{JHEP}
  {\bfseries 03} (2021) 257},
  [\href{https://arxiv.org/abs/2011.12373}{{\ttfamily 2011.12373}}].

\bibitem{ATLAS:2021ifb}
{\scshape ATLAS} collaboration, G.~Aad et~al., \emph{{Search for Higgs boson
  pair production in the two bottom quarks plus two photons final state in $pp$
  collisions at $\sqrt{s}$ = 13 TeV with the ATLAS detector}},
  \href{http://dx.doi.org/10.1103/PhysRevD.106.052001}{\emph{Phys. Rev. D}
  {\bfseries 106} (2022) 052001},
  [\href{https://arxiv.org/abs/2112.11876}{{\ttfamily 2112.11876}}].

\bibitem{CMS:2022hgz}
{\scshape CMS} collaboration, A.~M. Sirunyan et~al., \emph{{Search for
  nonresonant Higgs boson pair production in final state with two bottom quarks
  and two tau leptons in proton-proton collisions at $\sqrt{s}$ = 13 TeV}},
  \href{https://arxiv.org/abs/2206.09401}{{\ttfamily 2206.09401}}.

\bibitem{ATLAS:2022xzm}
{\scshape ATLAS} collaboration, G.~Aad et~al., \emph{{Search for resonant and
  non-resonant Higgs boson pair production in the $b\bar b\tau^+\tau^-$ decay
  channel using 13 TeV $pp$ collision data from the ATLAS detector}},
  \href{https://arxiv.org/abs/2209.10910}{{\ttfamily 2209.10910}}.

\bibitem{Grzadkowski:2010es}
B.~Grzadkowski, M.~Iskrzynski, M.~Misiak and J.~Rosiek, \emph{{Dimension-Six
  Terms in the Standard Model Lagrangian}},
  \href{http://dx.doi.org/10.1007/JHEP10(2010)085}{\emph{JHEP} {\bfseries 10}
  (2010) 085}, [\href{https://arxiv.org/abs/1008.4884}{{\ttfamily 1008.4884}}].

\bibitem{McCullough:2013rea}
M.~McCullough, \emph{{An Indirect Model-Dependent Probe of the Higgs
  Self-Coupling}},
  \href{http://dx.doi.org/10.1103/PhysRevD.90.015001}{\emph{Phys. Rev. D}
  {\bfseries 90} (2014) 015001},
  [\href{https://arxiv.org/abs/1312.3322}{{\ttfamily 1312.3322}}].

\bibitem{Degrassi:2016wml}
G.~Degrassi, P.~P. Giardino, F.~Maltoni and D.~Pagani, \emph{{Probing the Higgs
  self coupling via single Higgs production at the LHC}},
  \href{http://dx.doi.org/10.1007/JHEP12(2016)080}{\emph{JHEP} {\bfseries 12}
  (2016) 080}, [\href{https://arxiv.org/abs/1607.04251}{{\ttfamily
  1607.04251}}].

\bibitem{Gorbahn:2016uoy}
M.~Gorbahn and U.~Haisch, \emph{{Indirect probes of the trilinear Higgs
  coupling: $gg \to h$ and $h \to \gamma \gamma$}},
  \href{http://dx.doi.org/10.1007/JHEP10(2016)094}{\emph{JHEP} {\bfseries 10}
  (2016) 094}, [\href{https://arxiv.org/abs/1607.03773}{{\ttfamily
  1607.03773}}].

\bibitem{Kribs:2017znd}
G.~D. Kribs, A.~Maier, H.~Rzehak, M.~Spannowsky and P.~Waite,
  \emph{{Electroweak oblique parameters as a probe of the trilinear Higgs boson
  self-interaction}},
  \href{http://dx.doi.org/10.1103/PhysRevD.95.093004}{\emph{Phys. Rev. D}
  {\bfseries 95} (2017) 093004},
  [\href{https://arxiv.org/abs/1702.07678}{{\ttfamily 1702.07678}}].

\bibitem{Anisha:2022ctm}
Anisha, O.~Atkinson, A.~Bhardwaj, C.~Englert and P.~Stylianou, \emph{{Quartic
  Gauge-Higgs couplings: constraints and future directions}},
  \href{http://dx.doi.org/10.1007/JHEP10(2022)172}{\emph{JHEP} {\bfseries 10}
  (2022) 172}, [\href{https://arxiv.org/abs/2208.09334}{{\ttfamily
  2208.09334}}].

\bibitem{Agashe:2004rs}
K.~Agashe, R.~Contino and A.~Pomarol, \emph{{The Minimal composite Higgs
  model}}, \href{http://dx.doi.org/10.1016/j.nuclphysb.2005.04.035}{\emph{Nucl.
  Phys. B} {\bfseries 719} (2005) 165--187},
  [\href{https://arxiv.org/abs/hep-ph/0412089}{{\ttfamily hep-ph/0412089}}].

\bibitem{Contino:2006qr}
R.~Contino, L.~Da~Rold and A.~Pomarol, \emph{{Light custodians in natural
  composite Higgs models}},
  \href{http://dx.doi.org/10.1103/PhysRevD.75.055014}{\emph{Phys. Rev. D}
  {\bfseries 75} (2007) 055014},
  [\href{https://arxiv.org/abs/hep-ph/0612048}{{\ttfamily hep-ph/0612048}}].

\bibitem{Coleman:1969sm}
S.~R. Coleman, J.~Wess and B.~Zumino, \emph{{Structure of phenomenological
  Lagrangians. 1.}},
  \href{http://dx.doi.org/10.1103/PhysRev.177.2239}{\emph{Phys. Rev.}
  {\bfseries 177} (1969) 2239--2247}.

\bibitem{Callan:1969sn}
C.~G. Callan, Jr., S.~R. Coleman, J.~Wess and B.~Zumino, \emph{{Structure of
  phenomenological Lagrangians. 2.}},
  \href{http://dx.doi.org/10.1103/PhysRev.177.2247}{\emph{Phys. Rev.}
  {\bfseries 177} (1969) 2247--2250}.

\bibitem{Binoth:1996au}
T.~Binoth and J.~J. van~der Bij, \emph{{Influence of strongly coupled, hidden
  scalars on Higgs signals}},
  \href{http://dx.doi.org/10.1007/s002880050442}{\emph{Z. Phys. C} {\bfseries
  75} (1997) 17--25}, [\href{https://arxiv.org/abs/hep-ph/9608245}{{\ttfamily
  hep-ph/9608245}}].

\bibitem{Patt:2006fw}
B.~Patt and F.~Wilczek, \emph{{Higgs-field portal into hidden sectors}},
  \href{https://arxiv.org/abs/hep-ph/0605188}{{\ttfamily hep-ph/0605188}}.

\bibitem{Schabinger:2005ei}
R.~M. Schabinger and J.~D. Wells, \emph{{A Minimal spontaneously broken hidden
  sector and its impact on Higgs boson physics at the large hadron collider}},
  \href{http://dx.doi.org/10.1103/PhysRevD.72.093007}{\emph{Phys. Rev. D}
  {\bfseries 72} (2005) 093007},
  [\href{https://arxiv.org/abs/hep-ph/0509209}{{\ttfamily hep-ph/0509209}}].

\bibitem{Englert:2011yb}
C.~Englert, T.~Plehn, D.~Zerwas and P.~M. Zerwas, \emph{{Exploring the Higgs
  portal}}, \href{http://dx.doi.org/10.1016/j.physletb.2011.08.002}{\emph{Phys.
  Lett. B} {\bfseries 703} (2011) 298--305},
  [\href{https://arxiv.org/abs/1106.3097}{{\ttfamily 1106.3097}}].

\bibitem{Alonso:2015fsp}
R.~Alonso, E.~E. Jenkins and A.~V. Manohar, \emph{{A Geometric Formulation of
  Higgs Effective Field Theory: Measuring the Curvature of Scalar Field
  Space}}, \href{http://dx.doi.org/10.1016/j.physletb.2016.01.041}{\emph{Phys.
  Lett. B} {\bfseries 754} (2016) 335--342},
  [\href{https://arxiv.org/abs/1511.00724}{{\ttfamily 1511.00724}}].

\bibitem{Dawson:1984gx}
S.~Dawson, \emph{{The Effective W Approximation}},
  \href{http://dx.doi.org/10.1016/0550-3213(85)90038-0}{\emph{Nucl. Phys. B}
  {\bfseries 249} (1985) 42--60}.

\bibitem{Contino:2010mh}
R.~Contino, C.~Grojean, M.~Moretti, F.~Piccinini and R.~Rattazzi, \emph{{Strong
  Double Higgs Production at the LHC}},
  \href{http://dx.doi.org/10.1007/JHEP05(2010)089}{\emph{JHEP} {\bfseries 05}
  (2010) 089}, [\href{https://arxiv.org/abs/1002.1011}{{\ttfamily 1002.1011}}].

\bibitem{Hahn:1998yk}
T.~Hahn and M.~Perez-Victoria, \emph{{Automatized one loop calculations in
  four-dimensions and D-dimensions}},
  \href{http://dx.doi.org/10.1016/S0010-4655(98)00173-8}{\emph{Comput. Phys.
  Commun.} {\bfseries 118} (1999) 153--165},
  [\href{https://arxiv.org/abs/hep-ph/9807565}{{\ttfamily hep-ph/9807565}}].

\bibitem{Hahn:2000jm}
T.~Hahn, \emph{{Automatic loop calculations with FeynArts, FormCalc, and
  LoopTools}},
  \href{http://dx.doi.org/10.1016/S0920-5632(00)00848-3}{\emph{Nucl. Phys. B
  Proc. Suppl.} {\bfseries 89} (2000) 231--236},
  [\href{https://arxiv.org/abs/hep-ph/0005029}{{\ttfamily hep-ph/0005029}}].

\bibitem{Hahn:2000kx}
T.~Hahn, \emph{{Generating Feynman diagrams and amplitudes with FeynArts 3}},
  \href{http://dx.doi.org/10.1016/S0010-4655(01)00290-9}{\emph{Comput. Phys.
  Commun.} {\bfseries 140} (2001) 418--431},
  [\href{https://arxiv.org/abs/hep-ph/0012260}{{\ttfamily hep-ph/0012260}}].

\bibitem{DiLuzio:2017tfn}
L.~Di~Luzio, R.~Gr\"ober and M.~Spannowsky, \emph{{Maxi-sizing the trilinear
  Higgs self-coupling: how large could it be?}},
  \href{http://dx.doi.org/10.1140/epjc/s10052-017-5361-0}{\emph{Eur. Phys. J.
  C} {\bfseries 77} (2017) 788},
  [\href{https://arxiv.org/abs/1704.02311}{{\ttfamily 1704.02311}}].

\bibitem{Arco:2020ucn}
F.~Arco, S.~Heinemeyer and M.~J. Herrero, \emph{{Exploring sizable triple Higgs
  couplings in the 2HDM}},
  \href{http://dx.doi.org/10.1140/epjc/s10052-020-8406-8}{\emph{Eur. Phys. J.
  C} {\bfseries 80} (2020) 884},
  [\href{https://arxiv.org/abs/2005.10576}{{\ttfamily 2005.10576}}].

\bibitem{Arco:2022lai}
F.~Arco, S.~Heinemeyer, M.~M\"uhlleitner and K.~Radchenko, \emph{{Sensitivity
  to Triple Higgs Couplings via Di-Higgs Production in the 2HDM at the
  (HL-)LHC}},  \href{https://arxiv.org/abs/2212.11242}{{\ttfamily 2212.11242}}.

\bibitem{Cahn:1983ip}
R.~N. Cahn and S.~Dawson, \emph{{Production of Very Massive Higgs Bosons}},
  \href{http://dx.doi.org/10.1016/0370-2693(84)91180-8}{\emph{Phys. Lett. B}
  {\bfseries 136} (1984) 196}.

\bibitem{Rainwater:1998kj}
D.~L. Rainwater, D.~Zeppenfeld and K.~Hagiwara, \emph{{Searching for
  $H\to\tau^+\tau^-$ in weak boson fusion at the CERN LHC}},
  \href{http://dx.doi.org/10.1103/PhysRevD.59.014037}{\emph{Phys. Rev. D}
  {\bfseries 59} (1998) 014037},
  [\href{https://arxiv.org/abs/hep-ph/9808468}{{\ttfamily hep-ph/9808468}}].

\bibitem{Zeppenfeld:2000td}
D.~Zeppenfeld, R.~Kinnunen, A.~Nikitenko and E.~Richter-Was, \emph{{Measuring
  Higgs boson couplings at the CERN LHC}},
  \href{http://dx.doi.org/10.1103/PhysRevD.62.013009}{\emph{Phys. Rev. D}
  {\bfseries 62} (2000) 013009},
  [\href{https://arxiv.org/abs/hep-ph/0002036}{{\ttfamily hep-ph/0002036}}].

\bibitem{Figy:2003nv}
T.~Figy, C.~Oleari and D.~Zeppenfeld, \emph{{Next-to-leading order jet
  distributions for Higgs boson production via weak boson fusion}},
  \href{http://dx.doi.org/10.1103/PhysRevD.68.073005}{\emph{Phys. Rev. D}
  {\bfseries 68} (2003) 073005},
  [\href{https://arxiv.org/abs/hep-ph/0306109}{{\ttfamily hep-ph/0306109}}].

\bibitem{Figy:2008zd}
T.~Figy, \emph{{Next-to-leading order QCD corrections to light Higgs Pair
  production via vector boson fusion}},
  \href{http://dx.doi.org/10.1142/S0217732308028181}{\emph{Mod. Phys. Lett. A}
  {\bfseries 23} (2008) 1961--1973},
  [\href{https://arxiv.org/abs/0806.2200}{{\ttfamily 0806.2200}}].

\bibitem{Dreyer:2018qbw}
F.~A. Dreyer and A.~Karlberg, \emph{{Vector-Boson Fusion Higgs Pair Production
  at N$^3$LO}}, \href{http://dx.doi.org/10.1103/PhysRevD.98.114016}{\emph{Phys.
  Rev. D} {\bfseries 98} (2018) 114016},
  [\href{https://arxiv.org/abs/1811.07906}{{\ttfamily 1811.07906}}].

\bibitem{Arnold:2008rz}
K.~Arnold et~al., \emph{{VBFNLO: A Parton level Monte Carlo for processes with
  electroweak bosons}},
  \href{http://dx.doi.org/10.1016/j.cpc.2009.03.006}{\emph{Comput. Phys.
  Commun.} {\bfseries 180} (2009) 1661--1670},
  [\href{https://arxiv.org/abs/0811.4559}{{\ttfamily 0811.4559}}].

\bibitem{Baglio:2012np}
J.~Baglio, A.~Djouadi, R.~Gr\"ober, M.~M. M\"uhlleitner, J.~Quevillon and
  M.~Spira, \emph{{The measurement of the Higgs self-coupling at the LHC:
  theoretical status}},
  \href{http://dx.doi.org/10.1007/JHEP04(2013)151}{\emph{JHEP} {\bfseries 04}
  (2013) 151}, [\href{https://arxiv.org/abs/1212.5581}{{\ttfamily 1212.5581}}].

\bibitem{Alwall:2014hca}
J.~Alwall, R.~Frederix, S.~Frixione, V.~Hirschi, F.~Maltoni, O.~Mattelaer
  et~al., \emph{{The automated computation of tree-level and next-to-leading
  order differential cross sections, and their matching to parton shower
  simulations}}, \href{http://dx.doi.org/10.1007/JHEP07(2014)079}{\emph{JHEP}
  {\bfseries 07} (2014) 079},
  [\href{https://arxiv.org/abs/1405.0301}{{\ttfamily 1405.0301}}].

\bibitem{Frederix:2014hta}
R.~Frederix, S.~Frixione, V.~Hirschi, F.~Maltoni, O.~Mattelaer, P.~Torrielli
  et~al., \emph{{Higgs pair production at the LHC with NLO and parton-shower
  effects}},
  \href{http://dx.doi.org/10.1016/j.physletb.2014.03.026}{\emph{Phys. Lett. B}
  {\bfseries 732} (2014) 142--149},
  [\href{https://arxiv.org/abs/1401.7340}{{\ttfamily 1401.7340}}].

\bibitem{FCC:2018vvp}
{\scshape FCC} collaboration, A.~Abada et~al., \emph{{FCC-hh: The Hadron
  Collider}: {Future Circular Collider Conceptual Design Report Volume 3}},
  \href{http://dx.doi.org/10.1140/epjst/e2019-900087-0}{\emph{Eur. Phys. J. ST}
  {\bfseries 228} (2019) 755--1107}.

\bibitem{ATLAS:2018rnh}
{\scshape ATLAS} collaboration, M.~Aaboud et~al., \emph{{Search for pair
  production of Higgs bosons in the $b\bar{b}b\bar{b}$ final state using
  proton-proton collisions at $\sqrt{s} = 13$ TeV with the ATLAS detector}},
  \href{http://dx.doi.org/10.1007/JHEP01(2019)030}{\emph{JHEP} {\bfseries 01}
  (2019) 030}, [\href{https://arxiv.org/abs/1804.06174}{{\ttfamily
  1804.06174}}].

\bibitem{ATLAS:2020jgy}
{\scshape ATLAS} collaboration, G.~Aad et~al., \emph{{Search for the $HH
  \rightarrow b \bar{b} b \bar{b}$ process via vector-boson fusion production
  using proton-proton collisions at $\sqrt{s} = 13$ TeV with the ATLAS
  detector}}, \href{http://dx.doi.org/10.1007/JHEP07(2020)108}{\emph{JHEP}
  {\bfseries 07} (2020) 108},
  [\href{https://arxiv.org/abs/2001.05178}{{\ttfamily 2001.05178}}].

\bibitem{CMS:2022cpr}
{\scshape CMS} collaboration, A.~Tumasyan et~al., \emph{{Search for Higgs Boson
  Pair Production in the Four b Quark Final State in Proton-Proton Collisions
  at $\sqrt{s}$ = 13 TeV}},
  \href{http://dx.doi.org/10.1103/PhysRevLett.129.081802}{\emph{Phys. Rev.
  Lett.} {\bfseries 129} (2022) 081802},
  [\href{https://arxiv.org/abs/2202.09617}{{\ttfamily 2202.09617}}].

\bibitem{CMS:2022nmn}
{\scshape CMS} collaboration, A.~M. Sirunyan et~al., \emph{{Search for
  nonresonant pair production of highly energetic Higgs bosons decaying to
  bottom quarks}},  \href{https://arxiv.org/abs/2205.06667}{{\ttfamily
  2205.06667}}.

\bibitem{Bierlich:2022pfr}
C.~Bierlich et~al., \emph{{A comprehensive guide to the physics and usage of
  PYTHIA 8.3}},  \href{https://arxiv.org/abs/2203.11601}{{\ttfamily
  2203.11601}}.

\bibitem{Conte:2012fm}
E.~Conte, B.~Fuks and G.~Serret, \emph{{MadAnalysis 5, A User-Friendly
  Framework for Collider Phenomenology}},
  \href{http://dx.doi.org/10.1016/j.cpc.2012.09.009}{\emph{Comput. Phys.
  Commun.} {\bfseries 184} (2013) 222--256},
  [\href{https://arxiv.org/abs/1206.1599}{{\ttfamily 1206.1599}}].

\bibitem{Cacciari:2011ma}
M.~Cacciari, G.~P. Salam and G.~Soyez, \emph{{FastJet User Manual}},
  \href{http://dx.doi.org/10.1140/epjc/s10052-012-1896-2}{\emph{Eur. Phys. J.
  C} {\bfseries 72} (2012) 1896},
  [\href{https://arxiv.org/abs/1111.6097}{{\ttfamily 1111.6097}}].

\bibitem{Cacciari:2005hq}
M.~Cacciari and G.~P. Salam, \emph{{Dispelling the $N^{3}$ myth for the $k_t$
  jet-finder}},
  \href{http://dx.doi.org/10.1016/j.physletb.2006.08.037}{\emph{Phys. Lett. B}
  {\bfseries 641} (2006) 57--61},
  [\href{https://arxiv.org/abs/hep-ph/0512210}{{\ttfamily hep-ph/0512210}}].

\bibitem{Dolan:2013rja}
M.~J. Dolan, C.~Englert, N.~Greiner and M.~Spannowsky, \emph{{Further on up the
  road: $hhjj$ production at the LHC}},
  \href{http://dx.doi.org/10.1103/PhysRevLett.112.101802}{\emph{Phys. Rev.
  Lett.} {\bfseries 112} (2014) 101802},
  [\href{https://arxiv.org/abs/1310.1084}{{\ttfamily 1310.1084}}].

\bibitem{Dolan:2015zja}
M.~J. Dolan, C.~Englert, N.~Greiner, K.~Nordstrom and M.~Spannowsky,
  \emph{{$hhjj$ production at the LHC}},
  \href{http://dx.doi.org/10.1140/epjc/s10052-015-3622-3}{\emph{Eur. Phys. J.
  C} {\bfseries 75} (2015) 387},
  [\href{https://arxiv.org/abs/1506.08008}{{\ttfamily 1506.08008}}].

\bibitem{Bishara:2016kjn}
F.~Bishara, R.~Contino and J.~Rojo, \emph{{Higgs pair production in
  vector-boson fusion at the LHC and beyond}},
  \href{http://dx.doi.org/10.1140/epjc/s10052-017-5037-9}{\emph{Eur. Phys. J.
  C} {\bfseries 77} (2017) 481},
  [\href{https://arxiv.org/abs/1611.03860}{{\ttfamily 1611.03860}}].

\bibitem{ATLAS:2022vkf}
{\scshape ATLAS} collaboration, G.~Aad et~al., \emph{{A detailed map of Higgs
  boson interactions by the ATLAS experiment ten years after the discovery}},
  \href{http://dx.doi.org/10.1038/s41586-022-04893-w}{\emph{Nature} {\bfseries
  607} (2022) 52--59}, [\href{https://arxiv.org/abs/2207.00092}{{\ttfamily
  2207.00092}}].

\bibitem{Chacko:2017xpd}
Z.~Chacko, C.~Kilic, S.~Najjari and C.~B. Verhaaren, \emph{{Testing the Scalar
  Sector of the Twin Higgs Model at Colliders}},
  \href{http://dx.doi.org/10.1103/PhysRevD.97.055031}{\emph{Phys. Rev. D}
  {\bfseries 97} (2018) 055031},
  [\href{https://arxiv.org/abs/1711.05300}{{\ttfamily 1711.05300}}].

\bibitem{DiVita:2017vrr}
S.~Di~Vita, G.~Durieux, C.~Grojean, J.~Gu, Z.~Liu, G.~Panico et~al., \emph{{A
  global view on the Higgs self-coupling at lepton colliders}},
  \href{http://dx.doi.org/10.1007/JHEP02(2018)178}{\emph{JHEP} {\bfseries 02}
  (2018) 178}, [\href{https://arxiv.org/abs/1711.03978}{{\ttfamily
  1711.03978}}].

\bibitem{Li:2017daq}
B.~Li, Z.-L. Han and Y.~Liao, \emph{{Higgs production at future e$^{+}$e$^{-}$
  colliders in the Georgi-Machacek model}},
  \href{http://dx.doi.org/10.1007/JHEP02(2018)007}{\emph{JHEP} {\bfseries 02}
  (2018) 007}, [\href{https://arxiv.org/abs/1710.00184}{{\ttfamily
  1710.00184}}].

\bibitem{Abramowicz:2016zbo}
H.~Abramowicz et~al., \emph{{Higgs physics at the CLIC
  electron\textendash{}positron linear collider}},
  \href{http://dx.doi.org/10.1140/epjc/s10052-017-4968-5}{\emph{Eur. Phys. J.
  C} {\bfseries 77} (2017) 475},
  [\href{https://arxiv.org/abs/1608.07538}{{\ttfamily 1608.07538}}].

\bibitem{Domenech:2022uud}
D.~Domenech, M.~J. Herrero, R.~A. Morales and M.~Ramos, \emph{{Double Higgs
  boson production at TeV e+e- colliders with effective field theories:
  Sensitivity to BSM Higgs couplings}},
  \href{http://dx.doi.org/10.1103/PhysRevD.106.115027}{\emph{Phys. Rev. D}
  {\bfseries 106} (2022) 115027},
  [\href{https://arxiv.org/abs/2208.05452}{{\ttfamily 2208.05452}}].

\bibitem{Gonzalez-Lopez:2020lpd}
M.~Gonzalez-Lopez, M.~J. Herrero and P.~Martinez-Suarez, \emph{{Testing
  anomalous $H-W$ couplings and Higgs self-couplings via double and triple
  Higgs production at $e^+e^-$ colliders}},
  \href{http://dx.doi.org/10.1140/epjc/s10052-021-09048-1}{\emph{Eur. Phys. J.
  C} {\bfseries 81} (2021) 260},
  [\href{https://arxiv.org/abs/2011.13915}{{\ttfamily 2011.13915}}].

\bibitem{Roloff:2019crr}
{\scshape CLICdp} collaboration, P.~Roloff, U.~Schnoor, R.~Simoniello and
  B.~Xu, \emph{{Double Higgs boson production and Higgs self-coupling
  extraction at CLIC}},
  \href{http://dx.doi.org/10.1140/epjc/s10052-020-08567-7}{\emph{Eur. Phys. J.
  C} {\bfseries 80} (2020) 1010},
  [\href{https://arxiv.org/abs/1901.05897}{{\ttfamily 1901.05897}}].

\bibitem{deBlas:2019rxi}
J.~de~Blas et~al., \emph{{Higgs Boson Studies at Future Particle Colliders}},
  \href{http://dx.doi.org/10.1007/JHEP01(2020)139}{\emph{JHEP} {\bfseries 01}
  (2020) 139}, [\href{https://arxiv.org/abs/1905.03764}{{\ttfamily
  1905.03764}}].

\bibitem{Cepeda:2019klc}
M.~Cepeda et~al., \emph{{Report from Working Group 2}: {Higgs Physics at the
  HL-LHC and HE-LHC}},
  \href{http://dx.doi.org/10.23731/CYRM-2019-007.221}{\emph{CERN Yellow Rep.
  Monogr.} {\bfseries 7} (2019) 221--584},
  [\href{https://arxiv.org/abs/1902.00134}{{\ttfamily 1902.00134}}].

\bibitem{Contino:2016spe}
R.~Contino et~al., \emph{{Physics at a 100 TeV pp collider: Higgs and EW
  symmetry breaking studies}},
  \href{https://arxiv.org/abs/1606.09408}{{\ttfamily 1606.09408}}.

\bibitem{ILC:2019gyn}
{\scshape ILC} collaboration, H.~Aihara et~al., \emph{{The International Linear
  Collider. A Global Project}},
  \href{https://arxiv.org/abs/1901.09829}{{\ttfamily 1901.09829}}.

\bibitem{Baur:2002rb}
U.~Baur, T.~Plehn and D.~L. Rainwater, \emph{{Measuring the Higgs Boson Self
  Coupling at the LHC and Finite Top Mass Matrix Elements}},
  \href{http://dx.doi.org/10.1103/PhysRevLett.89.151801}{\emph{Phys. Rev.
  Lett.} {\bfseries 89} (2002) 151801},
  [\href{https://arxiv.org/abs/hep-ph/0206024}{{\ttfamily hep-ph/0206024}}].

\bibitem{Dolan:2012rv}
M.~J. Dolan, C.~Englert and M.~Spannowsky, \emph{{Higgs self-coupling
  measurements at the LHC}},
  \href{http://dx.doi.org/10.1007/JHEP10(2012)112}{\emph{JHEP} {\bfseries 10}
  (2012) 112}, [\href{https://arxiv.org/abs/1206.5001}{{\ttfamily 1206.5001}}].

\bibitem{Alonso:2021rac}
R.~Alonso and M.~West, \emph{{Roads to the Standard Model}},
  \href{http://dx.doi.org/10.1103/PhysRevD.105.096028}{\emph{Phys. Rev. D}
  {\bfseries 105} (2022) 096028},
  [\href{https://arxiv.org/abs/2109.13290}{{\ttfamily 2109.13290}}].

\bibitem{Gunion:1989ci}
J.~F. Gunion, R.~Vega and J.~Wudka, \emph{{Higgs triplets in the standard
  model}}, \href{http://dx.doi.org/10.1103/PhysRevD.42.1673}{\emph{Phys. Rev.
  D} {\bfseries 42} (1990) 1673--1691}.

\bibitem{Englert:2013wga}
C.~Englert, E.~Re and M.~Spannowsky, \emph{{Pinning down Higgs triplets at the
  LHC}}, \href{http://dx.doi.org/10.1103/PhysRevD.88.035024}{\emph{Phys. Rev.
  D} {\bfseries 88} (2013) 035024},
  [\href{https://arxiv.org/abs/1306.6228}{{\ttfamily 1306.6228}}].

\bibitem{CMS:2021wlt}
{\scshape CMS} collaboration, A.~M. Sirunyan et~al., \emph{{Search for charged
  Higgs bosons produced in vector boson fusion processes and decaying into
  vector boson pairs in proton\textendash{}proton collisions at $\sqrt{s} =$ 13
  TeV}}, \href{http://dx.doi.org/10.1140/epjc/s10052-021-09472-3}{\emph{Eur.
  Phys. J. C} {\bfseries 81} (2021) 723},
  [\href{https://arxiv.org/abs/2104.04762}{{\ttfamily 2104.04762}}].

\bibitem{Ismail:2020zoz}
A.~Ismail, H.~E. Logan and Y.~Wu, \emph{{Updated constraints on the
  Georgi-Machacek model from LHC Run 2}},
  \href{https://arxiv.org/abs/2003.02272}{{\ttfamily 2003.02272}}.

\bibitem{Egana-Ugrinovic:2015vgy}
D.~Egana-Ugrinovic and S.~Thomas, \emph{{Effective Theory of Higgs Sector
  Vacuum States}},  \href{https://arxiv.org/abs/1512.00144}{{\ttfamily
  1512.00144}}.

\bibitem{Banta:2021dek}
I.~Banta, T.~Cohen, N.~Craig, X.~Lu and D.~Sutherland, \emph{{Non-decoupling
  new particles}}, \href{http://dx.doi.org/10.1007/JHEP02(2022)029}{\emph{JHEP}
  {\bfseries 02} (2022) 029},
  [\href{https://arxiv.org/abs/2110.02967}{{\ttfamily 2110.02967}}].

\bibitem{Banta:2022rwg}
I.~Banta, \emph{{A strongly first-order electroweak phase transition from
  Loryons}}, \href{http://dx.doi.org/10.1007/JHEP06(2022)099}{\emph{JHEP}
  {\bfseries 06} (2022) 099},
  [\href{https://arxiv.org/abs/2202.04608}{{\ttfamily 2202.04608}}].

\bibitem{Ross:1973fp}
D.~A. Ross and J.~C. Taylor, \emph{{Renormalization of a unified theory of weak
  and electromagnetic interactions}},
  \href{http://dx.doi.org/10.1016/0550-3213(73)90505-1}{\emph{Nucl. Phys. B}
  {\bfseries 51} (1973) 125--144}.

\bibitem{Denner:2018opp}
A.~Denner, S.~Dittmaier and J.-N. Lang, \emph{{Renormalization of mixing
  angles}}, \href{http://dx.doi.org/10.1007/JHEP11(2018)104}{\emph{JHEP}
  {\bfseries 11} (2018) 104},
  [\href{https://arxiv.org/abs/1808.03466}{{\ttfamily 1808.03466}}].

\bibitem{Sirlin:1980nh}
A.~Sirlin, \emph{{Radiative Corrections in the SU(2)-L x U(1) Theory: A Simple
  Renormalization Framework}},
  \href{http://dx.doi.org/10.1103/PhysRevD.22.971}{\emph{Phys. Rev. D}
  {\bfseries 22} (1980) 971--981}.

\bibitem{Passarino:1978jh}
G.~Passarino and M.~J.~G. Veltman, \emph{{One Loop Corrections for $e^+ e^-$
  Annihilation Into $\mu^+ \mu^-$ in the Weinberg Model}},
  \href{http://dx.doi.org/10.1016/0550-3213(79)90234-7}{\emph{Nucl. Phys. B}
  {\bfseries 160} (1979) 151--207}.

\bibitem{Djouadi:2005gi}
A.~Djouadi, \emph{{The Anatomy of electro-weak symmetry breaking. I: The Higgs
  boson in the standard model}},
  \href{http://dx.doi.org/10.1016/j.physrep.2007.10.004}{\emph{Phys. Rept.}
  {\bfseries 457} (2008) 1--216},
  [\href{https://arxiv.org/abs/hep-ph/0503172}{{\ttfamily hep-ph/0503172}}].

\bibitem{Carmi:2012in}
D.~Carmi, A.~Falkowski, E.~Kuflik, T.~Volansky and J.~Zupan, \emph{{Higgs After
  the Discovery: A Status Report}},
  \href{http://dx.doi.org/10.1007/JHEP10(2012)196}{\emph{JHEP} {\bfseries 10}
  (2012) 196}, [\href{https://arxiv.org/abs/1207.1718}{{\ttfamily 1207.1718}}].

\bibitem{Contino:2003ve}
R.~Contino, Y.~Nomura and A.~Pomarol, \emph{{Higgs as a holographic
  pseudoGoldstone boson}},
  \href{http://dx.doi.org/10.1016/j.nuclphysb.2003.08.027}{\emph{Nucl. Phys. B}
  {\bfseries 671} (2003) 148--174},
  [\href{https://arxiv.org/abs/hep-ph/0306259}{{\ttfamily hep-ph/0306259}}].

\bibitem{Alonso:2016btr}
R.~Alonso, E.~E. Jenkins and A.~V. Manohar, \emph{{Sigma Models with Negative
  Curvature}},
  \href{http://dx.doi.org/10.1016/j.physletb.2016.03.032}{\emph{Phys. Lett. B}
  {\bfseries 756} (2016) 358--364},
  [\href{https://arxiv.org/abs/1602.00706}{{\ttfamily 1602.00706}}].

\bibitem{Alonso:2016oah}
R.~Alonso, E.~E. Jenkins and A.~V. Manohar, \emph{{Geometry of the Scalar
  Sector}}, \href{http://dx.doi.org/10.1007/JHEP08(2016)101}{\emph{JHEP}
  {\bfseries 08} (2016) 101},
  [\href{https://arxiv.org/abs/1605.03602}{{\ttfamily 1605.03602}}].

\bibitem{Ferretti:2014qta}
G.~Ferretti, \emph{{UV Completions of Partial Compositeness: The Case for a
  SU(4) Gauge Group}},
  \href{http://dx.doi.org/10.1007/JHEP06(2014)142}{\emph{JHEP} {\bfseries 06}
  (2014) 142}, [\href{https://arxiv.org/abs/1404.7137}{{\ttfamily 1404.7137}}].

\bibitem{Contino:2010rs}
R.~Contino, \emph{{The Higgs as a Composite Nambu-Goldstone Boson}},  in
  \emph{{Theoretical Advanced Study Institute in Elementary Particle Physics}:
  {Physics of the Large and the Small}}, pp.~235--306, 2011.
\newblock \href{https://arxiv.org/abs/1005.4269}{{\ttfamily 1005.4269}}.
\newblock \href{http://dx.doi.org/10.1142/9789814327183_0005}{DOI}.

\bibitem{Golterman:2015zwa}
M.~Golterman and Y.~Shamir, \emph{{Top quark induced effective potential in a
  composite Higgs model}},
  \href{http://dx.doi.org/10.1103/PhysRevD.91.094506}{\emph{Phys. Rev. D}
  {\bfseries 91} (2015) 094506},
  [\href{https://arxiv.org/abs/1502.00390}{{\ttfamily 1502.00390}}].

\bibitem{DelDebbio:2017ini}
L.~Del~Debbio, C.~Englert and R.~Zwicky, \emph{{A UV Complete Compositeness
  Scenario: LHC Constraints Meet The Lattice}},
  \href{http://dx.doi.org/10.1007/JHEP08(2017)142}{\emph{JHEP} {\bfseries 08}
  (2017) 142}, [\href{https://arxiv.org/abs/1703.06064}{{\ttfamily
  1703.06064}}].

\bibitem{Ayyar:2018glg}
V.~Ayyar, T.~DeGrand, D.~C. Hackett, W.~I. Jay, E.~T. Neil, Y.~Shamir et~al.,
  \emph{{Partial compositeness and baryon matrix elements on the lattice}},
  \href{http://dx.doi.org/10.1103/PhysRevD.99.094502}{\emph{Phys. Rev. D}
  {\bfseries 99} (2019) 094502},
  [\href{https://arxiv.org/abs/1812.02727}{{\ttfamily 1812.02727}}].

\bibitem{DelDebbio:2022qgu}
L.~Del~Debbio, A.~Lupo, M.~Panero and N.~Tantalo, \emph{{Multi-representation
  dynamics of SU(4) composite Higgs models: chiral limit and spectral
  reconstructions}},
  \href{http://dx.doi.org/10.1140/epjc/s10052-023-11363-8}{\emph{Eur. Phys. J.
  C} {\bfseries 83} (2023) 220},
  [\href{https://arxiv.org/abs/2211.09581}{{\ttfamily 2211.09581}}].

\bibitem{Grober:2010yv}
R.~Grober and M.~Muhlleitner, \emph{{Composite Higgs Boson Pair Production at
  the LHC}}, \href{http://dx.doi.org/10.1007/JHEP06(2011)020}{\emph{JHEP}
  {\bfseries 06} (2011) 020},
  [\href{https://arxiv.org/abs/1012.1562}{{\ttfamily 1012.1562}}].

\bibitem{Durieux:2021riy}
G.~Durieux, M.~McCullough and E.~Salvioni, \emph{{Gegenbauer Goldstones}},
  \href{http://dx.doi.org/10.1007/JHEP01(2022)076}{\emph{JHEP} {\bfseries 01}
  (2022) 076}, [\href{https://arxiv.org/abs/2110.06941}{{\ttfamily
  2110.06941}}].

\bibitem{alma9938889293902959}
J.~A. Wolf, \emph{Spaces of constant curvature}.
\newblock AMS Chelsea Pub., Providence, R.I, 6th ed.~ed., 2011.

\bibitem{Bekenstein:1992pj}
J.~D. Bekenstein, \emph{{The Relation between physical and gravitational
  geometry}}, \href{http://dx.doi.org/10.1103/PhysRevD.48.3641}{\emph{Phys.
  Rev. D} {\bfseries 48} (1993) 3641--3647},
  [\href{https://arxiv.org/abs/gr-qc/9211017}{{\ttfamily gr-qc/9211017}}].

\bibitem{Bruggisser:2022ofg}
S.~Bruggisser, B.~von Harling, O.~Matsedonskyi and G.~Servant, \emph{{Dilaton
  at the LHC: Complementary Probe of Composite Higgs}},
  \href{https://arxiv.org/abs/2212.00056}{{\ttfamily 2212.00056}}.

\bibitem{Goldberger:2007zk}
W.~D. Goldberger, B.~Grinstein and W.~Skiba, \emph{{Distinguishing the Higgs
  boson from the dilaton at the Large Hadron Collider}},
  \href{http://dx.doi.org/10.1103/PhysRevLett.100.111802}{\emph{Phys. Rev.
  Lett.} {\bfseries 100} (2008) 111802},
  [\href{https://arxiv.org/abs/0708.1463}{{\ttfamily 0708.1463}}].

\bibitem{Rattazzi:2000hs}
R.~Rattazzi and A.~Zaffaroni, \emph{{Comments on the holographic picture of the
  Randall-Sundrum model}},
  \href{http://dx.doi.org/10.1088/1126-6708/2001/04/021}{\emph{JHEP} {\bfseries
  04} (2001) 021}, [\href{https://arxiv.org/abs/hep-th/0012248}{{\ttfamily
  hep-th/0012248}}].

\bibitem{Komargodski:2011vj}
Z.~Komargodski and A.~Schwimmer, \emph{{On Renormalization Group Flows in Four
  Dimensions}}, \href{http://dx.doi.org/10.1007/JHEP12(2011)099}{\emph{JHEP}
  {\bfseries 12} (2011) 099},
  [\href{https://arxiv.org/abs/1107.3987}{{\ttfamily 1107.3987}}].

\bibitem{Dolan:2012ac}
M.~J. Dolan, C.~Englert and M.~Spannowsky, \emph{{New Physics in LHC Higgs
  boson pair production}},
  \href{http://dx.doi.org/10.1103/PhysRevD.87.055002}{\emph{Phys. Rev. D}
  {\bfseries 87} (2013) 055002},
  [\href{https://arxiv.org/abs/1210.8166}{{\ttfamily 1210.8166}}].

\bibitem{Herrero:2022krh}
M.~J. Herrero and R.~A. Morales, \emph{{One-loop corrections for WW to HH in
  Higgs EFT with the electroweak chiral Lagrangian}},
  \href{http://dx.doi.org/10.1103/PhysRevD.106.073008}{\emph{Phys. Rev. D}
  {\bfseries 106} (2022) 073008},
  [\href{https://arxiv.org/abs/2208.05900}{{\ttfamily 2208.05900}}].

\bibitem{Herrero:2021iqt}
M.~J. Herrero and R.~A. Morales, \emph{{One-loop renormalization of vector
  boson scattering with the electroweak chiral Lagrangian in covariant
  gauges}}, \href{http://dx.doi.org/10.1103/PhysRevD.104.075013}{\emph{Phys.
  Rev. D} {\bfseries 104} (2021) 075013},
  [\href{https://arxiv.org/abs/2107.07890}{{\ttfamily 2107.07890}}].

\bibitem{Delgado:2013hxa}
R.~L. Delgado, A.~Dobado and F.~J. Llanes-Estrada, \emph{{One-loop $W_LW_L$ and
  $Z_LZ_L$ scattering from the electroweak Chiral Lagrangian with a light
  Higgs-like scalar}},
  \href{http://dx.doi.org/10.1007/JHEP02(2014)121}{\emph{JHEP} {\bfseries 02}
  (2014) 121}, [\href{https://arxiv.org/abs/1311.5993}{{\ttfamily 1311.5993}}].

\bibitem{Gavela:2014uta}
M.~B. Gavela, K.~Kanshin, P.~A.~N. Machado and S.~Saa, \emph{{On the
  renormalization of the electroweak chiral Lagrangian with a Higgs}},
  \href{http://dx.doi.org/10.1007/JHEP03(2015)043}{\emph{JHEP} {\bfseries 03}
  (2015) 043}, [\href{https://arxiv.org/abs/1409.1571}{{\ttfamily 1409.1571}}].

\bibitem{Asiain:2021lch}
I.~n. Asi\'ain, D.~Espriu and F.~Mescia, \emph{{Introducing tools to test Higgs
  boson interactions via WW scattering: One-loop calculations and
  renormalization in the Higgs effective field theory}},
  \href{http://dx.doi.org/10.1103/PhysRevD.105.015009}{\emph{Phys. Rev. D}
  {\bfseries 105} (2022) 015009},
  [\href{https://arxiv.org/abs/2109.02673}{{\ttfamily 2109.02673}}].

\bibitem{Gomez-Ambrosio:2022why}
R.~G\'omez-Ambrosio, F.~J. Llanes-Estrada, A.~Salas-Bern\'ardez and J.~J.
  Sanz-Cillero, \emph{{SMEFT is falsifiable through multi-Higgs measurements
  (even in the absence of new light particles)}},
  \href{http://dx.doi.org/10.1088/1572-9494/ace95e}{\emph{Commun. Theor. Phys.}
  {\bfseries 75} (2023) 095202},
  [\href{https://arxiv.org/abs/2207.09848}{{\ttfamily 2207.09848}}].

\bibitem{Gomez-Ambrosio:2022qsi}
R.~G\'omez-Ambrosio, F.~J. Llanes-Estrada, A.~Salas-Bern\'ardez and J.~J.
  Sanz-Cillero, \emph{{Distinguishing electroweak EFTs with
  WLWL\textrightarrow{}n\texttimes{}h}},
  \href{http://dx.doi.org/10.1103/PhysRevD.106.053004}{\emph{Phys. Rev. D}
  {\bfseries 106} (2022) 053004},
  [\href{https://arxiv.org/abs/2204.01763}{{\ttfamily 2204.01763}}].

\bibitem{Dedes:2017zog}
A.~Dedes, W.~Materkowska, M.~Paraskevas, J.~Rosiek and K.~Suxho, \emph{{Feynman
  rules for the Standard Model Effective Field Theory in R$_\xi$ -gauges}},
  \href{http://dx.doi.org/10.1007/JHEP06(2017)143}{\emph{JHEP} {\bfseries 06}
  (2017) 143}, [\href{https://arxiv.org/abs/1704.03888}{{\ttfamily
  1704.03888}}].

\bibitem{ATLAS:2023dbw}
{\scshape ATLAS} collaboration, G.~Aad et~al., \emph{{Measurement and
  interpretation of same-sign $W$ boson pair production in association with two
  jets in $pp$ collisions at $\sqrt{s} = 13$ TeV with the ATLAS detector}},
  {\emph{ATLAS-CONF-2023-023} (2023) }.

\bibitem{ATLAS:2022hnn}
{\scshape ATLAS} collaboration, G.~Aad et~al., \emph{{Search for
  pair-production of vector-like quarks in pp collision events at s=13 TeV with
  at least one leptonically decaying Z boson and a third-generation quark with
  the ATLAS detector}},
  \href{http://dx.doi.org/10.1016/j.physletb.2023.138019}{\emph{Phys. Lett. B}
  {\bfseries 843} (2023) 138019},
  [\href{https://arxiv.org/abs/2210.15413}{{\ttfamily 2210.15413}}].

\bibitem{ATLAS:2020fry}
{\scshape ATLAS} collaboration, G.~Aad et~al., \emph{{Search for heavy diboson
  resonances in semileptonic final states in pp collisions at $\sqrt{s}=13$ TeV
  with the ATLAS detector}},
  \href{http://dx.doi.org/10.1140/epjc/s10052-020-08554-y}{\emph{Eur. Phys. J.
  C} {\bfseries 80} (2020) 1165},
  [\href{https://arxiv.org/abs/2004.14636}{{\ttfamily 2004.14636}}].

\bibitem{Arco:2023sac}
F.~Arco, D.~Domenech, M.~J. Herrero and R.~A. Morales, \emph{{Non-decoupling
  effects from heavy Higgs bosons by matching 2HDM to HEFT amplitudes}},
  \href{https://arxiv.org/abs/2307.15693}{{\ttfamily 2307.15693}}.

\bibitem{Hartling:2014zca}
K.~Hartling, K.~Kumar and H.~E. Logan, \emph{{The decoupling limit in the
  Georgi-Machacek model}},
  \href{http://dx.doi.org/10.1103/PhysRevD.90.015007}{\emph{Phys. Rev. D}
  {\bfseries 90} (2014) 015007},
  [\href{https://arxiv.org/abs/1404.2640}{{\ttfamily 1404.2640}}].

\end{thebibliography}\endgroup
\end{document}